\documentclass[fleqn,usenatbib]{mnras}
\usepackage[T1]{fontenc}
\usepackage{ae,aecompl}
 \usepackage{epsfig}
 \usepackage{amsmath}
 \usepackage{amssymb}
 \usepackage{natbib}
 \usepackage{multirow}
 \usepackage{graphicx}
 \usepackage{color}
 \usepackage{epstopdf}
 \usepackage{float}
 \usepackage{multirow}
 \usepackage{tabularx}

\title[Galaxy Accretion Fractions]{Accreted or Not Accreted? The Fraction of Accreted Mass in Galaxies from Simulations and Observations}

\author[R.-S. Remus and D. A. Forbes]{
Rhea-Silvia Remus$^{1}$\thanks{E-mail: rhea@usm.lmu.de} and 
Duncan A. Forbes$^{2}$
\\
$^{1}$Universit\"ats-Sternwarte M\"unchen, Fakult\"at f\"ur Physik, LMU M\"unchen, Scheinerstr.\ 1, D-81679 M\"unchen, Germany\\
$^{2}$Centre for Astrophysics and Supercomputing, Swinburne University of Technology, Hawthorn VIC 3122, Australia}

\date{Accepted XXX. Received YYY; in original form ZZZ}

\pubyear{2021}
\begin{document}
\label{firstpage}
\pagerange{\pageref{firstpage}--\pageref{lastpage}}
\maketitle

\begin{abstract}
In the two-phase scenario of galaxy formation, a galaxy's stellar mass growth is first dominated by in-situ star formation, and subsequently by accretion. We analyse the radial distribution of the accreted stellar mass in $\sim$500 galaxies from the hydrodynamical cosmological simulation Magneticum. Generally, we find good agreement with other simulations in that higher mass galaxies have larger accreted fractions, but we predict higher accretion fractions for low-mass galaxies. Based on the radial distribution of the accreted and in-situ components, we define 6 galaxy classes, from completely accretion dominated to completely in-situ dominated, and measure the transition radii between in-situ and accretion-dominated regions for galaxies that have such a transition. About 70\% of our galaxies have one transition radius. However, we also find about 10\% of the galaxies to be accretion dominated everywhere, and about 13\% to have two transition radii, with the centre and the outskirts both being accretion dominated. We show that these classes are strongly correlated with the galaxy merger histories, especially with the mergers' cold gas fractions. We find high total in-situ (low accretion) fractions to be associated with smaller, lower mass galaxies, lower central dark matter fractions, and larger transition radii. Finally, we show that the dips in observed surface brightness profiles seen in many early-type galaxies do \textit{not} correspond to the transition from in-situ to accretion-dominated regions, and any inferred mass fractions are not indicative of the true accreted mass. Instead, these dips contain information about the galaxies' dry minor merger assembly history.
\end{abstract}

\begin{keywords}
galaxies: structure -- galaxies: evolution -- galaxies: formation -- methods: numerical -- methods: observational
\end{keywords}


\section{Introduction}
In the two-phase scenario of galaxy formation \citep[e.g.,][]{oser:2010,pillepich:2014}, galaxies undergo two main phases of growth: first the in-situ, and subsequently the ex-situ (or accretion) growth phase. In the former phase, stars are formed within the primary galaxy. In the latter phase, mass growth occurs through accretion of satellite galaxies. 
A pioneering work in this area was presented by \citet{oser:2010}, using a set of zoom simulations of massive galaxies, who found that the in-situ phase occurs between redshifts 6 and 2, giving rise to a `galaxy core' of size $\sim$2~kpc, followed by the accretion dominated growth at lower redshifts. They also find higher mass galaxies to have higher fractions of accreted mass, and that the accreted mass is more often deposited in the galaxy outer halo regions. This has been subsequently supported by parameter studies using binary merger simulations of different mass ratios, demonstrating that larger mass ratios for host and satellite galaxies are more likely to lead to a deposition of the accreted stellar mass at larger radii, while small merger mass ratios usually lead to full mixture of the accreted and in-situ formed stars \citep[e.g.,][]{hilz:2012,karademir:2019}. However, the picture is not that clear, as the orbital configurations of the mergers have been shown to influence the radius of mass deposition for satellite galaxies especially for mergers of larger mass ratios, with circular orbits leading to mass depositions at larger radii than radial merger orbits \citep[e.g.,][]{amorisco:2017,karademir:2019}.

This two-phase scenario has been further refined over the last decade with increasingly sophisticated, full cosmological simulations. Those focusing on predictions for the accreted stellar component of early-type galaxies include the dark matter particle tagging approach of \citet{cooper:2013,cooper:2015BCG} and, more recently, hydrodynamical cosmological simulations like Illustris \citep{pillepich:2014,rodriguez:2016}, EAGLE \citep{davidson:2020}, and Illustris-TNG \citep{tacchella:2019,pulsoni:2020}.

In particular, \citet{cooper:2013} modeled 1872 central galaxies in the mass range $10.7 < \log M_* < 11.4$  (i.e., $11.5 <\log M_\mathrm{200} < 14.0$) using a semi-analytic method to tag dark matter particles, not including a full treatment of gas physics in the simulation itself. Being mainly massive galaxies, the sample is dominated by early-type galaxies. They found that accretion leads to a break, or change, in the slope of the stellar mass surface density profile at the radius where the accreted material starts to dominate over that formed in-situ. They fit S\'ersic profiles to the in-situ and accreted stars separately in surface density space, finding that the resulting double S\'ersic profiles provided a good overall fit. 
They showed that the fraction of accreted material approaches 100\% for the most massive early-type galaxies, with more accretion dominated galaxies having shallower density profiles with little, or no, obvious transition in the overall profile. This study was further expanded for the very high mass end ($\log M_\mathrm{200} \sim 14$) by \citet{cooper:2015BCG}. They found double S\'ersic profiles to be a good fit to the stellar mass surface density profiles, with the inner component having S\'ersic indices of $n \sim 4$ and the outer component having $n \sim 1$ (similar to observational results for BCGs, e.g., \citet{2007MNRAS.378.1575S,kluge:2019}). However, these inner and outer components were found to corresponded to the relaxed and unrelaxed accreted stars rather than the in-situ and accreted stars.

Results from the fully hydrodynamical cosmological simulations all confirm the idea of the two-phase scenario of galaxy formation, and the general trend for higher mass galaxies to have larger amounts of accreted stars and shallower radial stellar density profiles \citep{pillepich:2014,rodriguez:2016,tacchella:2019,davidson:2020}. They also generally agree that the transition radius between accretion-dominated and in-situ-dominated stars is generally smaller for more massive galaxies. However, they show very different results when it comes to the details of the accreted versus in-situ components of galaxies, caused by the different subgrid models describing the star formation and feedback processes \citep[see][for a recent review]{vogelsberger:2020}. 

For example, \citet{rodriguez:2016} find for the old Illustris simulations that {\it all} galaxies reveal a clear `transition' radius for which the in-situ and accreted components contribute equally (i.e., 50:50), while \citet{tacchella:2019} reported for the new Illustris-TNG simulations that their most massive galaxies can be dominated by accreted stars at all radii. In addition, \citet{tacchella:2019} find much higher accreted mass fractions at a given stellar mass for the new Illustris-TNG simulations than \citet{rodriguez:2016} found for the old Illustris simulatios. 
Using the EAGLE simulation, \citet{davidson:2020} found similar ex-situ fractions as a function of stellar mass as \citet{tacchella:2019}. They also confirmed previous studies findings that the majority of accreted material is deposited in the outer regions, while for the most massive galaxies the accreted mass can dominate over in-situ material in the inner regions. However, the mass-size relations found for the galaxies from these two simulations are different, clearly showing that the details of galaxy formation still strongly differ between the different simulations.

More recently, these studies were also broadened to study the radial kinematic profiles of galaxies as possible tracers for the transition radii from in-situ to accretion dominated parts of galaxies (e.g., \citet{schulze:2020} using the Magneticum simulations and \citet{pulsoni:2020} using the Illustris-TNG simulations). Both studies find that the shape of the kinematic profile (i.e., $v/\sigma$) does not, in general, trace the transition between in-situ and ex-situ dominated galaxy regions. \citet{schulze:2020} showed that the kinematic profiles only for a special subset of galaxies that only experienced very small mergers since $z\sim1$ can be used as tracer for the transition radius, \citet{pulsoni:2020} split their galaxies into four classes of stellar mass density profiles based on the variation of the in-situ and ex-situ fractions with radius. We comment further on their findings in the main sections of this work. 

Observationally, the in-situ and accreted components of galaxies are much harder to tackle. Deep imaging studies of massive early-type galaxies (ETGs) found that around 3/4 of their galaxies reveal evidence for substructures in their stellar halos (e.g., 
\citet{1992AJ....104.1039S,1992MNRAS.254..723F,2009AJ....138.1417T,2015MNRAS.446..120D,kluge:2019}). Such substructures, in the form of shells, plumes, envelopes etc, likely represent the debris of accreted satellite galaxies. 
After stacking a large sample of 42,000 luminous red galaxies (with $\log M_*> 11$), \citet{2011ApJ...731...89T} found that within about $8R_\mathrm{e}$ their surface brightness profiles could be well represented by a single Serisc profile, but beyond that an extra component was required. A transition at similar radii have been reported in the globular cluster systems of massive ETGs (e.g., \citet{2018MNRAS.479.4760F}). 
\citet{dsouza:2014} fit 45,500 galaxies (avoiding edge-on disks) in several stellar mass bins, finding good fits to a double S\'ersic. 
There have also been claims that the surface brightness profiles of elliptical galaxies are better represented by three components 
\citep{2013ApJ...766...47H}, with radii of $<1R_\mathrm{e}$, $\sim2.5R_\mathrm{e}$, and $\sim10R_\mathrm{e}$, all with S\'ersic values of $n\sim{}$1--2. 
Thus, a range of radii from much smaller than $1R_\mathrm{e}$ to 8--10$R_\mathrm{e}$, have been reported in the literature as key transition radii. 
Very few studies have quantified the transition radius between different galaxy components and the mass associated with each component. This has, however, been attempted by \citet{spavone:2017} and \citet{spavone:2020}, using deep imaging of ETGs from the VEGAS survey. Fitting double S\'ersic functions to the galaxy surface brightness profiles and deriving transition radii from this they inferred outer halo mass fractions, finding evidence for higher mass fractions in the outer component for more massive galaxies. 

We note that several late-type galaxies have been studied in order to measure their outer halo light, e.g., 
the deep imaging of \citet{2016ApJ...830...62M} using the Dragonfly camera. This study revealed a large range in the fraction of halo light in late-type galaxies beyond 5 disk scale lengths from $\sim10\%$ to $<0.01\%$. However, it is not clear in how far this represents the accreted component of the disk galaxies as this outer light could also come from extended thick disk components.

In this work, we investigate the in-situ and accreted components using galaxies from the hydrodynamical cosmological simulation Magneticum. We compare their total and radial accretion properties to results from other simulations as well as observations of massive early-type galaxies, especially addressing the question in how far the transition radii from accreted to in-situ components can be inferred from S\'ersic fits to the radial surface density profiles.
In Sec.~\ref{sec:2} we present the simulations and the details of our classification of in-situ and accreted. Results from the simulations are presented in Sec.~\ref{sec:3}. This includes the classification of the radial density profiles into 6 classes (\ref{sec:31}), probing their assembly history (\ref{sec:32}), a comparison with other simulations (\ref{sec:33}), and a study of the correlation of the in-situ/accreted fractions with various galaxy properties (\ref{sec:34}) and the transition radii (\ref{sec:35}). Sec.~\ref{sec:4} provides a comparison between the 2D density profiles from simulations with observations and the question of possible recovery of the accreted fractions from the projected profiles. Finally we present our summary and conclusions in Sec.~\ref{sec:5}. 

\section{The Magneticum Pathfinder Simulations}\label{sec:2}
We use the Magneticum Pathfinder\footnote{www.magneticum.org} simulations (Dolag et al. 2021, in prep.),
which are a set of cosmological hydrodynamical SPH-simulations of several boxes with volumes ranging from 
$(2688~\mathrm{Mpc}/h)^3$ to $(48~\mathrm{Mpc}/h)^3$ and different resolutions, with the lowest having 
$m_\mathrm{Gas} = 2.6\times 10^9 M_\odot/h$ and the currently highest having $m_\mathrm{Gas} = 7.3\times 10^6 M_\odot/h$. 
Each gas particle can spawn up to four stellar particles during its lifetime, and as such the average mass of a stellar particle
is $1/4$th of the gas particle for each resolution.
A WMAP7 $\Lambda$CDM cosmology \citep{komatsu:2011} is adapted throughout all simulations,
with $\sigma_8 =0.809$, $h = 0.704$, $\Omega_\Lambda = 0.728$, $\Omega_\mathrm{M} = 0.272$, $\Omega_\mathrm{B} = 0.0451$,
and an initial slope for the power spectrum of $n_\mathrm{s} = 0.963$.

All simulations are performed with a version of GADGET-3 that includes various updates in the formulation of SPH \citep{dolag:2004,dolag:2005,donnert:2013,beck:2015}
as well as in the sub-grid physics, especially with respect to the star formation and metal enrichment descriptions \citep{tornatore:2004,tornatore:2007,wiersma:2009}
and the black hole feedback \citep{fabjan:2010,hirschmann:2014}. For more details on the physics included in the Magneticum Pathfinder simulations 
we refer the reader to \citet{hirschmann:2014,teklu:2015} and \citet{dolag:2017}.
Structures are identified using a modified version of SUBFIND \citep{springel:2001,dolag:2009}.

As shown in previous works, the Magneticum Pathfinder simulations successfully reproduce many observational results over a broad range of masses, from galaxy clusters down to field galaxies. Most relevant for the work presented here, they capture the evolution and properties of black holes (BHs) and AGN \citep{hirschmann:2014,steinborn:2015,steinborn:2016}, and the angular momentum, kinematic, and dynamical 
properties of galaxies at low and high redshifts \citep{teklu:2015,remus:2017,schulze:2018,teklu:2018,vandesande:2019}.
Especially relevant for the work presented in this paper, the Magneticum spheroidal and disk galaxies successfully match the observed size-mass relation up to redshifts of $z=2$ \citep{remus:2017,schulze:2018}.

\subsection{High resolution simulation}
As we are focusing on the internal properties of galaxies in this work, we use the currently largest volume of Magneticum with the highest resolution level available. This box has a size of $(48~\mathrm{Mpc}/h)^{3}$. It initially contains a total of $2\times576^{3}$ (dark matter and gas) particles. 
The mass resolution for the dark matter, gas, and stellar particles is $m_\mathrm{DM} = 3.6\times10^{7} M_{\odot}/h$, 
$m_\mathrm{Gas} = 7.3\times10^{6} M_{\odot}/h$, and $m_\mathrm{*} \simeq 2\times10^{6} M_{\odot}/h$, respectively, with a 
softening of $\epsilon_\mathrm{DM} = \epsilon_\mathrm{Gas} = 1.4~\mathrm{kpc}/h$ for dark matter and gas particles, and 
$\epsilon_\mathrm{*} = 0.7~\mathrm{kpc}/h$ for stellar particles. 

We choose a lower stellar mass limit of $M_\mathrm{*} \geq 2\times10^{10} M_{\odot}$ to ensure sufficient resolution for radial density profile fits, and additionally limit the sample to central galaxies to ensure a proper treatment of the in-situ/ex-situ classification. The highest mass galaxies in this simulation are $M_\mathrm{*}\sim10^{12} M_{\odot}$. With these restrictions, we select 511 galaxies, which include 4 galaxies that are brightest cluster galaxies (BCGs), and 43 galaxies which are brightest group galaxies. 

\subsection{Definition of Accreted and In-situ Stars}
We are interested in the accreted (ex-situ) and the in-situ components of the galaxies. Therefore, we have to trace all stars that are part of a galaxy at z=0 back to their formation redshift. If the star is born inside the main-branch progenitor of the galaxy, it is considered to be formed ``in-situ''. If the star was born outside the virial radius of the main-branch progenitor of the galaxy, and only later in its life accreted onto that  galaxy, then it is considered to be ``accreted'' independent of whether it was accreted smoothly or as part of another galaxy. If the star particle is born inside the virial radius of the main-branch progenitor but in the wake of a gas-rich merger, we still consider the star to be formed ``in-situ'', as otherwise all stars would be accreted since ultimately all gas has been accreted onto the galaxy. These stars have been handled differently in the literature, however, for the sake of a clean classification with respect to the accreted fraction we use the classification described above. This gives us the smallest possible fraction of accreted stars, and the largest possible fraction of in-situ formed stars.
\begin{figure*}
    \includegraphics[width=0.85\textwidth]{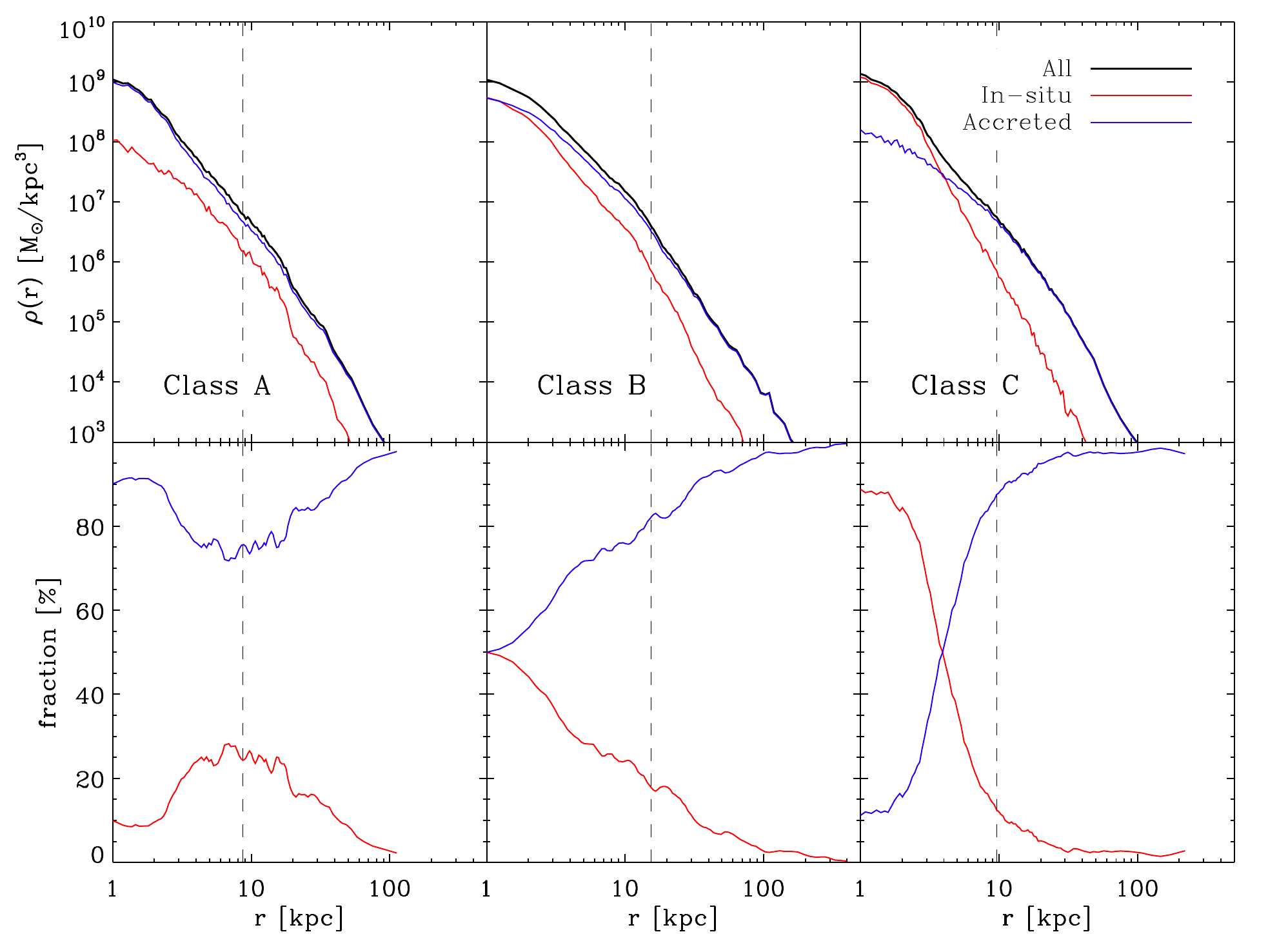}
    \includegraphics[width=0.84\textwidth]{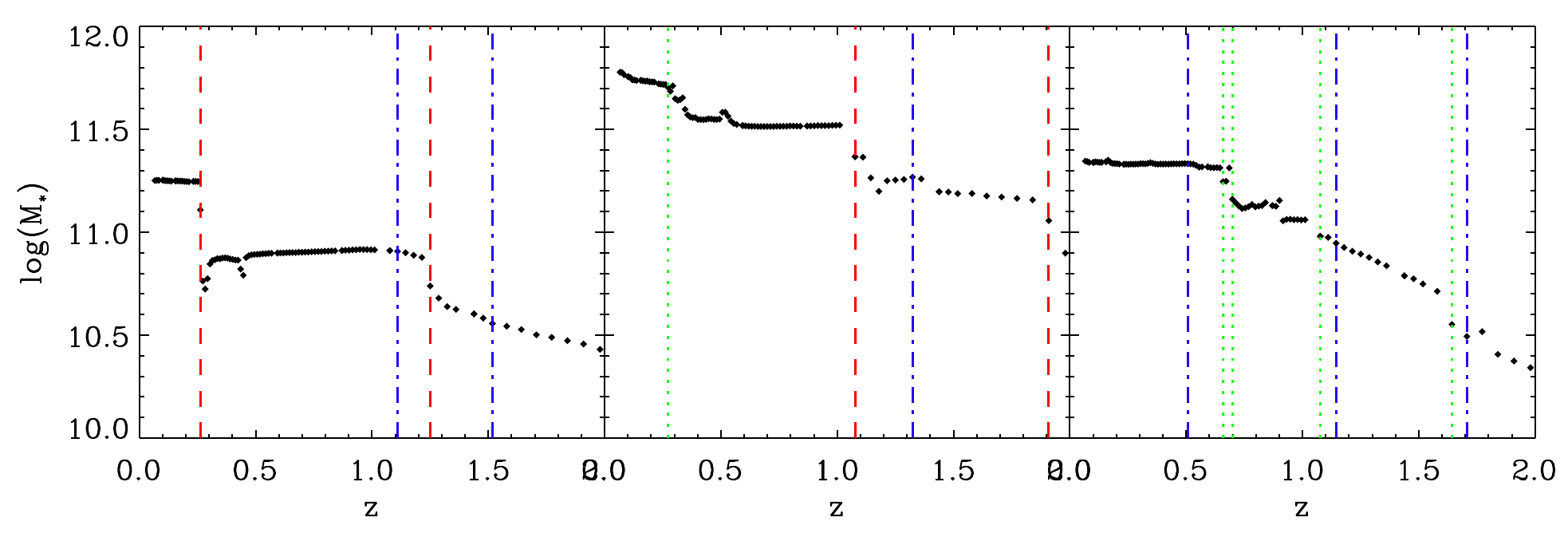}
    \caption{Examples for three of the six different in-situ/accreted profile classes, from left to right: class~A (extremely accretion dominated), class~B (accretion dominated), and class~C (classic). The vertical black dashed lines in the upper panels indicate the half mass radius. \textit{Upper panels:} Radial stellar density profiles for all stars (black lines), in-situ formed stars (red lines) and accreted stars (blue lines). \textit{Middle panels:} Relative mass fractions of the in-situ (red) and accreted (blue) subcomponents. 
     \textit{Bottom panels:} The assembly history of the stellar mass of the example galaxies. Dashed red lines show major mergers (mass ratios of 1:1 to 3:1), green lines show minor mergers (mass ratios of 3:1 to 10:1), and blue lines show mini mergers (mass ratios below 10:1).
}
    \label{fig:example_big_classes}
\end{figure*}
\begin{figure*}
    \includegraphics[width=0.85\textwidth]{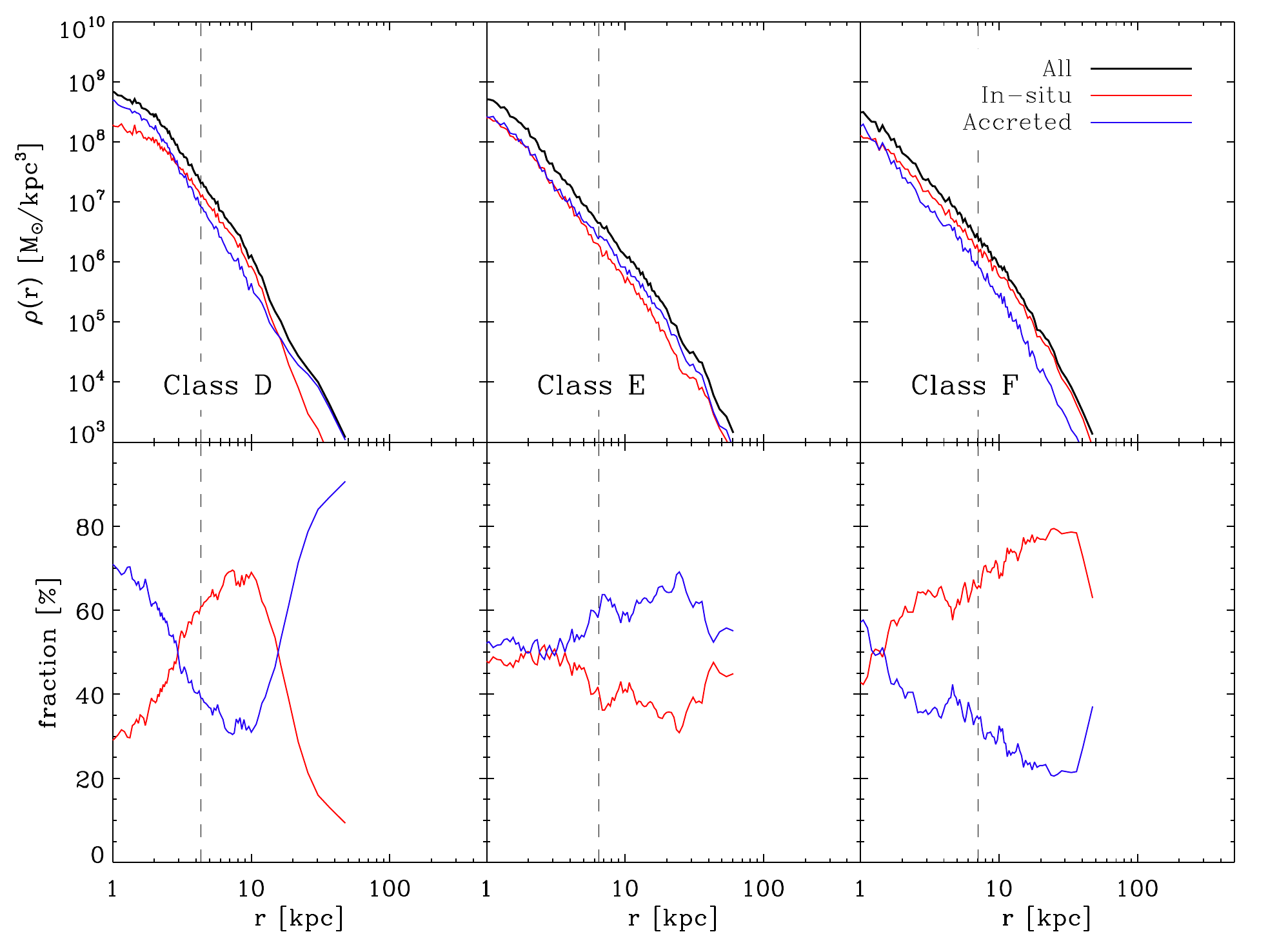}
    \includegraphics[width=0.84\textwidth]{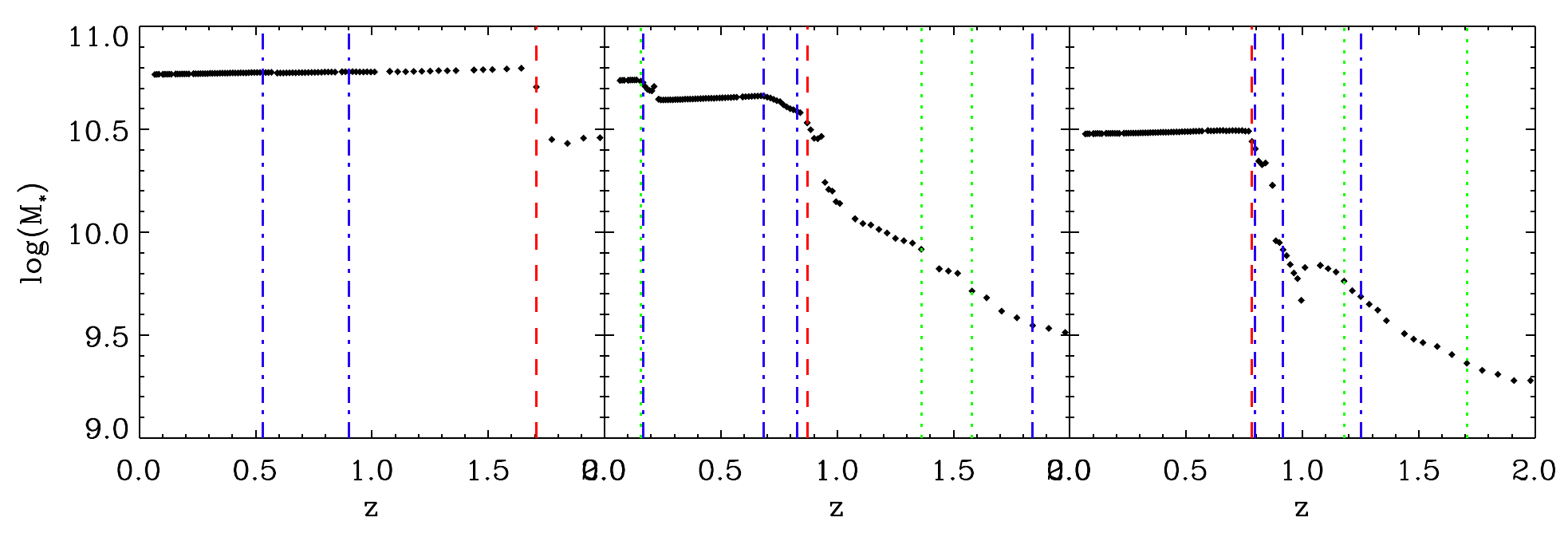}
    \caption{Same as Fig.~\ref{fig:example_big_classes} but for the other three profile classes, from left to right: class~D (double cross-over), class~E (balanced), and class~F (in-situ dominated).
        }
    \label{fig:example_minor_classes}
\end{figure*}

\subsection{Galaxy Classification}
The sample of 511 Magneticum galaxies includes all galaxy types. However, in the second part of this study, we restrict our investigation to spheroidal (or early-type) galaxies. They are selected using the $b$-value
$$b = \log_{10}\left(\frac{j_*}{\mathrm{kpc}~\mathrm{km/s}}\right)-\frac{2}{3}\log_{10}\left(\frac{M_*}{\mathrm{M_\odot}}\right),$$
which effectively gives a galaxies' position in the $M_*$--$j_*$ plane as discussed by \citet{teklu:2015}. At $z=0$, galaxies with a $b$-value of $b \leq -4.73$ are classified as spheroidals, while galaxies with $b \geq -4.35$ are classified as disks \citep{teklu:2017}. Galaxies with $b$-values in between these limits have intermediate properties, that is they include S0 galaxies and disk galaxies with large bulges, but also a small 
number of ongoing-merger and interacting galaxies. On this basis our sample includes 154 spheroidal, 105 disk, and 252 intermediate galaxies.
This is the same classification that has been used by \citet{teklu:2017}, \citet{schulze:2018}, and \citet{schulze:2020}.

\section{Accreted and In-situ Formed Stars in Magneticum Galaxies}\label{sec:3}
\subsection{Radial Stellar Mass Density Profiles}\label{sec:31}
We calculate the radial stellar mass density profiles for the Magneticum galaxies using equal particle bins with at least 200 particles per bin, in spherical shells around the galaxy center. For the in-situ and the accreted components, the same radial bins are used as for the total profile to ensure a direct radial comparability of the two components. To accommodate for the softening, we exclude the inner 1.4~kpc, but we do not use an outer radial limit.
\begin{figure*}
    \includegraphics[width=0.85\textwidth]{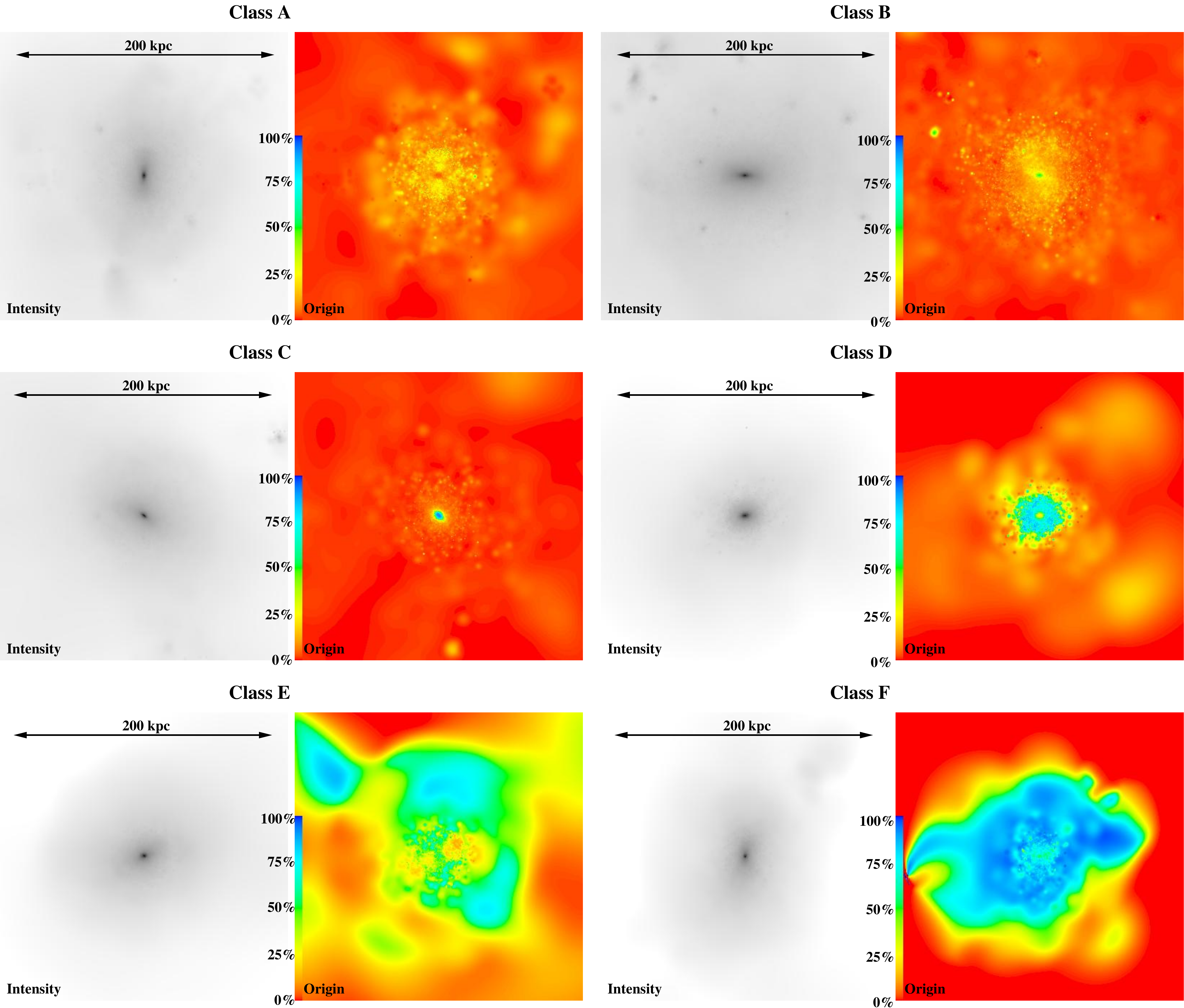}
    \caption{Random 2D projected views of the example Magneticum galaxies for each of the six profile classes. For each galaxy, a box with a length of 200~kpc centred on the galaxy is shown, with the left plot showing the intensity map derived from all stars in the galaxy and the right plot showing the origin of the stars colour coded according to the in-situ/accreted fraction (with blue colours showing 100\% in-situ fractions and red colours showing 100\% accreted fraction).
    From left to right, top to bottom, the shown galaxies are from class~A (upper left), class~B (upper right), class~C (central left), class~D (central right), class~E (bottom left), and class~F (bottom right). As can be clearly seen, the amount of in-situ stars increases from the upper left class (A) to the lower right class (F).
}
    \label{fig:showplots}
\end{figure*}

Inspecting the 3D radial stellar density profiles (in $M_\odot/\mathrm{kpc}^3$) of the Magneticum galaxies, we find six different classes based on their in-situ/accreted behaviour. Examples for each class are shown in Fig.~\ref{fig:example_big_classes} and Fig.~\ref{fig:example_minor_classes}, and Fig.~\ref{fig:showplots}, and described in the following:
\begin{itemize}
\item {\bf Class A:} Extremely accretion dominated profiles. For these galaxies, the accreted stellar component is always dominant, even in the inner regions (see left panels of Fig.~\ref{fig:example_big_classes} and upper left panels of Fig.~\ref{fig:showplots}). About 7\% of all galaxies show this kind of behaviour (see Tab.~\ref{tab:classes}). 
Such galaxies have no clear transition radius between in-situ and accreted stellar mass.
\item {\bf Class B:} Accretion dominated profiles. For these galaxies, the fraction of in-situ and accreted stars near the galaxy center is equal, but for all larger radii the accreted fraction dominates (see middle panels of Fig.~\ref{fig:example_big_classes} and upper right panels of Fig.~\ref{fig:showplots}). This is a rare class, with only about 2\% of all Magneticum galaxies in this class  (see Tab.~\ref{tab:classes}). It could also be interpreted as an extreme case of class~A, but we here study it as a separate class. The transition radius for these galaxies is very small, and is not a real transition in all cases as the in-situ component does not necessarily dominate in the center but are sometimes simply equal amounts of in-situ and accreted.
\item {\bf Class C:} Classic profiles. The inner regions of these galaxies are dominated by in-situ formed stars, while in the outskirts the accreted stellar component is dominant (see right panels of Fig.~\ref{fig:example_big_classes} and central left panels of Fig.~\ref{fig:showplots}). This is by far the most common class of profiles, with 72\% of all Magneticum galaxies showing this behaviour (see Tab.~\ref{tab:classes}). This is also the behaviour found most commonly in previous work for example by \citet{cooper:2010,rodriguez:2016,pulsoni:2020}.
These galaxies have a clear transition radius from in-situ to accretion dominated.
\item {\bf Class D:} Double cross-over profiles. These galaxies have a large accreted fraction dominating in the inner \textit{and} the outer regions, with their intermediate-radii regions dominated by in-situ formed stars (see left panels of Fig.~\ref{fig:example_minor_classes} and central right panels of Fig.~\ref{fig:showplots}). 13\% of all Magneticum galaxies fall in this category (see Tab.~\ref{tab:classes}), making this the second most common profile type. Given its nature, these profiles have \textit{two} transition radii.
\item {\bf Class E:} Balanced profiles. A small fraction ($\approx 4\%$, see Tab.~\ref{tab:classes}) of all Magneticum galaxies reveal profiles for which the in-situ and accreted contributions are nearly equal over a large radial range (see middle panels of Fig.~\ref{fig:example_minor_classes} and bottom left panels of Fig.~\ref{fig:showplots}). For these profiles, we usually find a transition radius at very large radii, however, even if the outer parts are slightly dominated by accreted stars, the fraction of accreted stars usually stays below 60\%.
\item {\bf Class F:} In-situ dominated profiles. Galaxies in this class have radial density profiles that are always dominated by in-situ formed stars at all radii, even at their outskirts (see right panels of Fig.~\ref{fig:example_minor_classes} and bottom right panels of Fig.~\ref{fig:showplots}). Only 2.7\% of all Magneticum galaxies show this behaviour, with the in-situ fraction always larger than the accreted fraction (Tab.~\ref{tab:classes}). 
As for class~A galaxies, there is no transition radius for these galaxies. 
\end{itemize}
\begin{table}
\centering
\begin{tabularx}{.9\columnwidth}{l||XX|XX|XX}
\hline
Class    & \multicolumn{2}{c|}{All} & \multicolumn{2}{c|}{Spheroidals} & \multicolumn{2}{c}{Disks}\\
                 &  N   & \%            &       N       &       \%      &       N       &       \%      \\
\hline\hline
Class A &       36      &       7.1             &       15      &       9.7             &       1       &       0.95            \\
Class B &       9       &       1.8             &       6       &       3.9             &       1       &       0.95            \\
Class C &       367     &       71.8    &       100     &       64.9    &       78      &       74.3    \\
Class D         &       66      &       12.9    &       24      &       15.6    &       18      &       17.1    \\
Class E         &       19      &       3.7             &       7       &       4.6             &       2       &       1.9             \\
Class F         &       14      &       2.7             &       2       &       1.3             &       5       &       4.8             \\
\hline
\end{tabularx}
\caption{Numbers and percentage for the six different profile classes for all 511 galaxies from the Magneticum simulation used in this work, 
and for those galaxies that are classified as spheroidals (154 galaxies) or disks (105 galaxies).
}
\label{tab:classes}
\end{table}

Recently, a similar analysis of radial in-situ and accreted profiles has been presented by \citet{pulsoni:2020} using the Illustris-TNG simulations. Differently to our six classes of 3D mass density profiles, they reported four classes based on 2D mass surface density profiles:
Their class~1 (20\% of the sample) galaxies are in-situ dominated at all radii, equivalent to our class~F galaxies, albeit our class~F only covers 2.7\% of the galaxies. Class~2 is their most common profile (57\%), and is equivalent to our class~C, albeit we have more galaxies of class~C (72\%). This is also the kind of in-situ/accreted profiles that have solely been reported for Illustris galaxies \citep{rodriguez:2016}. Their class~3 (15\%) is closest to our double cross-over profiles (class~D), and with 13\% of our galaxies being class~D the numbers are very similar between the two simulations. Their class~4 galaxies are accretion dominated and their least common profile at 8\%. In our work such profiles were divided into class~A (extremely accretion dominated) and~B (accretion dominated), representing a total of about 9\% of galaxies, again in good agreement with each others. We also classified $\sim4\%$ of our galaxies to be `balanced' with similar in-situ and accreted components, a class that does not appear in the classifications by \citet{pulsoni:2020}. Bearing in mind that the relative proportions of each profile class will depend on the range of galaxy types and masses in each simulation, the main differences are that we find more classic profiles (72\% vs 57\%) and fewer in-situ dominated profiles (2.7\% vs 20\%), clearly highlighting the intrinsic differences between the simulations.

\subsection{Assembly History}\label{sec:32}
\begin{figure*}
    \includegraphics[width=0.9\columnwidth]{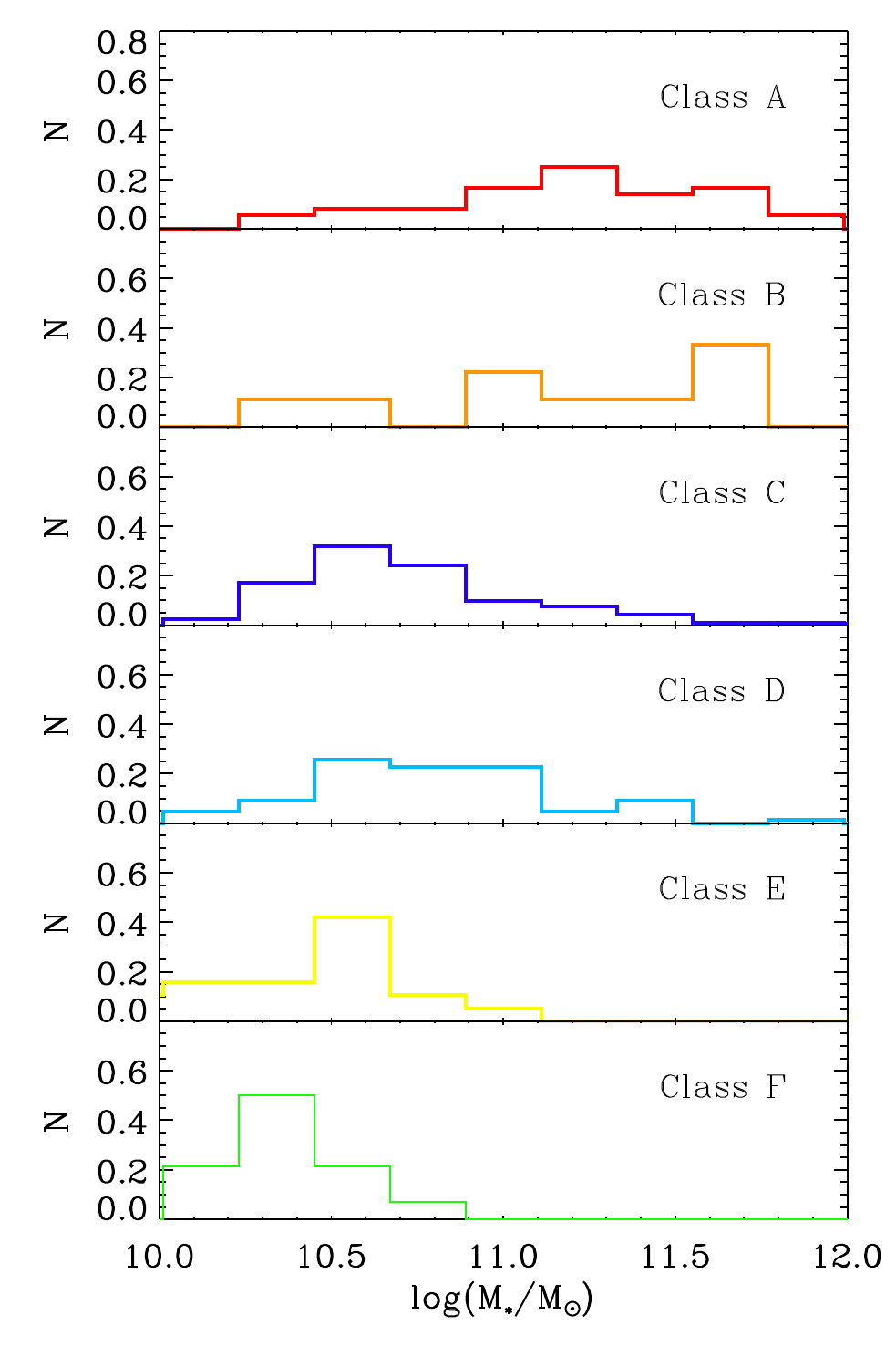}
    \includegraphics[width=0.9\columnwidth]{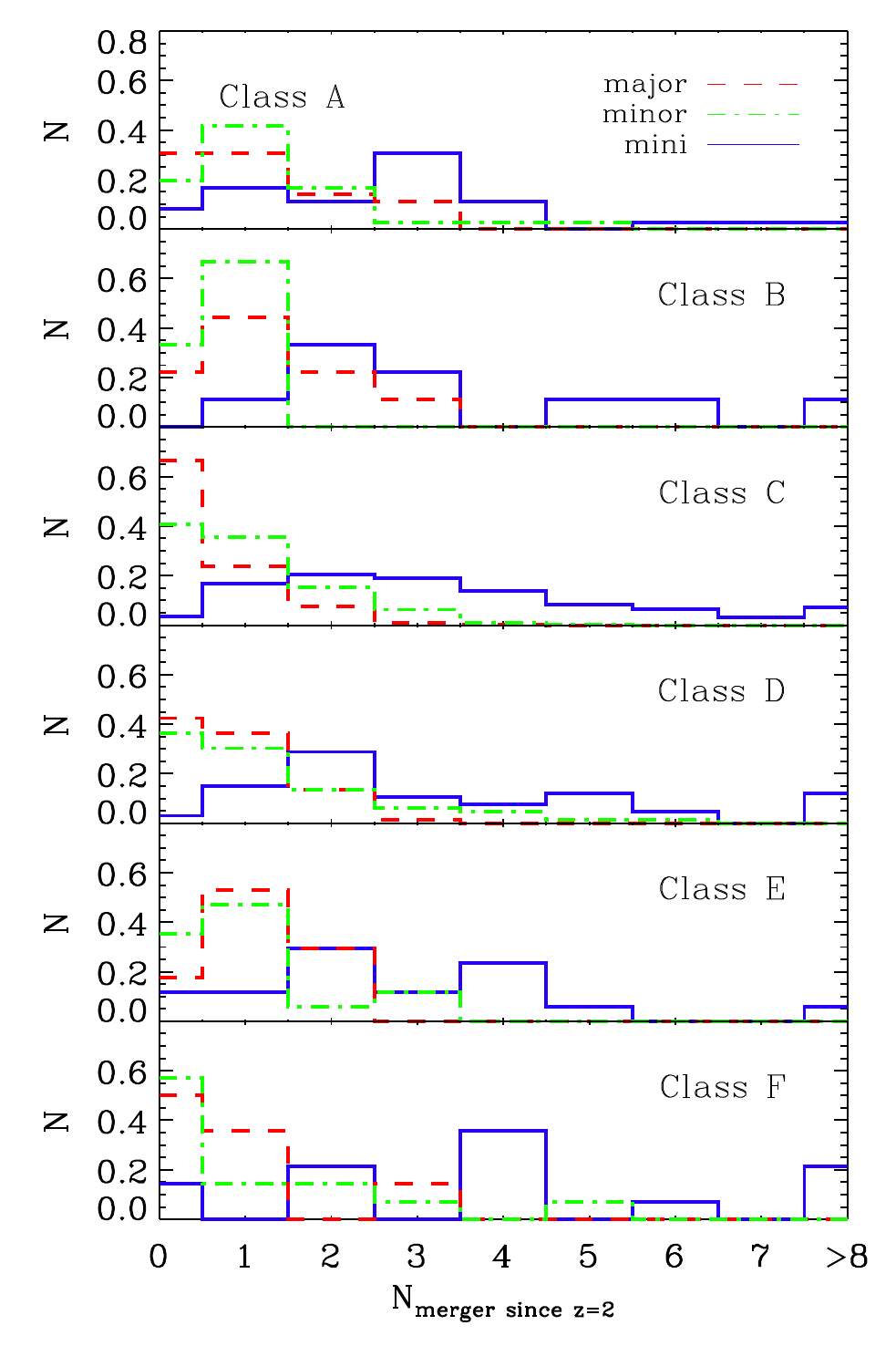}
    \caption{Profile class dependence on stellar mass and merger frequency. \textit{Left panel:} Histogram of the stellar mass distribution, colour coded by profile class. Low mass galaxies are dominated by classes C, E and F, while classes A and B are common for high mass galaxies. 
    \textit{Right panels:} Histogram showing the frequency of major (red), minor (green), and mini (blue) mergers since $z = 2$ for each class. Major mergers are least common for galaxies of class~C, where nearly 60\% of the galaxies have had no major merger since z=2.  
	}
    \label{fig:mass_classes}
\end{figure*}

One of the first steps to understand the origin of the different profile classes is to test whether they correlate with the stellar mass of the galaxy.
In the left panel of Fig.~\ref{fig:mass_classes} we show the normalised distribution of the different profile classes as a function of stellar mass. We find a clear trend that galaxies of classes~E and F, where the in-situ component is 50\% or larger, are always galaxies with relatively low stellar masses, while galaxies of classes A and B, which are everywhere accretion dominated, usually reside at the high mass end.
This mass trend is expected, as more massive galaxies have generally accreted more mass than low mass galaxies. Similar trends were seen by \citet{pulsoni:2020} in their study. 
Interestingly, the two most common classes of galaxies, namely classes~C and D, are most likely to be found in the middle mass range of our galaxy sample, with the galaxies having stellar masses of $10.5<\log(M_*)<11$. This indicates that it is not the frequency of mergers, but the type of merger that is crucial in establishing the differences.  
\begin{figure}
    \includegraphics[width=0.9\columnwidth]{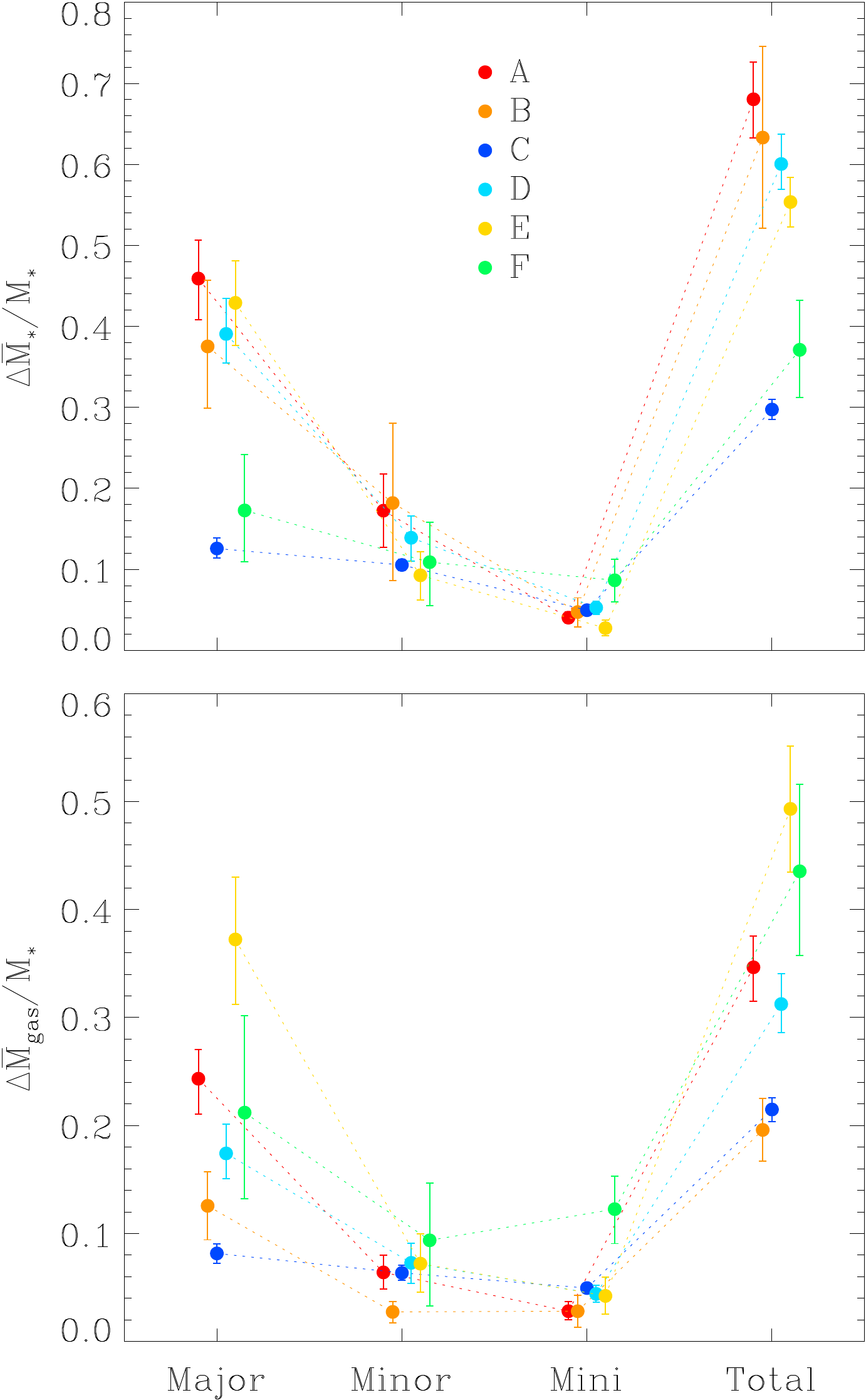}
        \caption{Mass fraction added via major, minor, mini mergers and all mergers together since $z=2$ for the six different profile classes. Classes A/B/C/D/E/F are given as red/orange/blue/cyan/yellow/green filled circles, respectively. \textit{Top panel:} Fraction of stellar mass accreted through mergers of different types. \textit{Bottom panel:} Fraction of gas mass accreted through mergers of different types. Gas mass includes hot and cold gas.  For the most massive galaxies a significant fraction of the gas is in a hot (and thus not star forming) phase.
	}
    \label{fig:accretion_hist}
\end{figure}

To understand this in more detail, we study the assembly history of all galaxies with respect to their main formation branch from $z=2$ to $z=0$. 
We distinguish three different kind of mergers: 
\begin{itemize}
\item Major mergers with mass ratios of 1:1 to 3:1.
\item Minor mergers with mass ratios of 3:1 to 10:1.
\item Mini mergers with mass ratios below 10:1.
\end{itemize}
For the mini mergers, there is no lower mass limit in general, however, due to the resolution limit of the simulation, the smallest mergers that we can resolve here are on the order of 100:1 in mass ratio. Below this limit, we do not call accretion events mergers, even for those few cases where this would still be resolvable, for the sake of completeness in numbers.

Major mergers are known to have strong impacts on the main progenitor galaxy at all radii, however, this is not in general the case for minor and mini mergers.
While minor mergers, especially in the mass range around 5:1, can still strongly influence the mass distribution of the progenitor galaxies even at their centres \citep[e.g.,][]{hilz:2013,karademir:2019}, mini mergers mostly contribute to the outer halos of galaxies and only play a role for the central evolution of the host galaxy for radial infall orbits or head-on collisions \citep{karademir:2019}. 
Since we focus on radial ranges beyond the half-mass radius, all types of mergers can play a role in establishing the different profile classes.

Examples of the assembly history for the six classes are given in the lower panels of Fig.~\ref{fig:example_big_classes} and Fig.~\ref{fig:example_minor_classes}, with the history always belonging to the galaxy for which the radial density profiles are shown in the upper panels of the same figures, and the intensity maps are shown in the according panels of Fig.~\ref{fig:showplots}.
The redshifts of  past merger events are marked as red/green/blue dashed lines for major/minor/mini mergers, respectively. 
These six examples show that all galaxies, but one, experience major mergers, with the galaxy that experiences no major merger being of class~C. 
In the case of the example galaxies from classes A and B (the accretion dominated profile classes), both show {\it two} major merger events since $z=2$. Interestingly, the galaxy from class~F, which is dominated by in-situ stars at all radii, experiences a major merger at a redshift of about $z=1$.

More quantitatively, in the right panel of Fig.~\ref{fig:mass_classes} we show for each profile class histograms of the frequency of major (red dashed lines), minor (green dash-dotted lines), and mini mergers (blue solid line) since redshift $z=2$ ($\sim10~\mathrm{Gyr}$ in look-back time). Furthermore,  Fig.~\ref{fig:accretion_hist} shows the amount of stellar mass relative to the present-day stellar mass that was accreted through the mergers of different mass ratios
and through all mergers for each accretion class in the upper panel, while the lower panel shows the fraction of gas relative to the present-day stellar mass accreted through these mergers for the different accretion classes.

In general, about half of all Magneticum galaxies have experienced at least one major merger since $z=2$.
However, when considering individual profile classes, we find large differences. Most strikingly, major mergers generally play a significant role in the formation of galaxies from accretion classes A, B, D, and E, while they only play a minor role in the evolution of galaxies from accretion classes C and F. 
This reflects what could already be seen from the example cases, but more statistically we find the following accretion history patterns for our six accretion profile classes:
\begin{itemize}
\item \textbf{Class A:} The accretion history of this ``overmerged'' class of galaxies is not surprisingly completely dominated by merger events, with a total of nearly
70\% of the present-day stellar mass being accreted (see upper panel of Fig.~\ref{fig:accretion_hist}). Surprisingly, these mergers are not necessarily dry.
Especially the major mergers contribute an average of 25\% of the total stellar mass in gas, but the shear amount of accreted stellar mass is enough to dominate the final galaxy at all radii. 
Additionally, most of these galaxies are rather massive, and as such a larger amount of the gas will be present as a hot gas halo and not participate in the star formation process.
Galaxies of this profile class are rather common for the spheroidals, but only one of these is found among the disk galaxies\footnote{This is a very special case in which the accretion occurs along a plane along the galaxies disk-plane, similar to what was discussed for mini mergers by \citet{karademir:2019}}.
\item \textbf{Class B:} Galaxies from this accretion class have a similar accretion history to galaxies of class~A, with more than 60\% of their stellar mass being accreted.
The main difference here is that the mergers were all gas poor, resulting in the lowest amount of gas accreted since $z=2$ in the whole sample (see lower panel of Fig.~\ref{fig:accretion_hist}).
\item \textbf{Class C:} The ``classic'' profile shows a significantly different behaviour from all the others, namely 2/3 of the galaxies in this class have never experienced a major merger since $z=2$ (see right panel of Fig.~\ref{fig:mass_classes}). 
Even those class~C galaxies with major mergers only get about 10\% of their mass from this pathway, implying that the major mergers happen rather early for this class (see upper panel of Fig.~\ref{fig:accretion_hist}). 
In general, galaxies of this class only accrete about 30\% of their stellar mass through mergers, which is the lowest fraction found for all classes.
Furthermore, the mass accreted through the major and minor mergers is approximately equal, and the relative contribution of the mini mergers is rather large
for galaxies of this group. This clearly shows the importance of minor and mini mergers in the assembly history of class~C galaxies. In addition,
we find that the mergers that a galaxy of class~C experiences, are usually rather dry, and contribute only about 20\% of the total stellar mass in gas (see lower panel of Fig.~\ref{fig:accretion_hist}).
This also explains the dominance of the accreted material at large radii and the dominance of the in-situ components in the center, as the minor and mini merger, especially when dry,
often do not reach the central parts of the host galaxy at all but rather deposit their mass at large radii \citep{purcell:2007,amorisco:2017,karademir:2019}. 
\item \textbf{Class D:}
Major mergers are important for half of the galaxies in this profile class, but those mergers are relatively dry (gas-poor). 
The host galaxy, on the other hand, is relatively wet (gas-rich) at the time of merging, and through the merger the gas is moved outwards into a ring-like structure, where the star formation occurs. 
Our simulation does not have the resolution to confirm this, but this may be one possible way to form an (old) bulge inside a gas-rich galaxy. In addition, these galaxies are equally common for both disks and spheroidals (seen Tab.~\ref{tab:classes}).
\item \textbf{Class E:} The mass accretion history of galaxies from this class is dominated by a single merger which is either a major merger (60\%) or a massive minor merger (see right panels of Fig.~\ref{fig:mass_classes}).
These mergers were gas-rich, causing a starburst after accretion, which effectively leads to this special profile case where the in-situ and accreted radial fractions are identical over a broad radial range.
As known from classical binary merger simulations \citet[e.g.,][]{1991Natur.354..210H}, 
these mergers usually result in a spheroidal galaxy as long as the merger is not in-plane or has a very high gas fraction \citep{springel:2005}.
This is reflected in the low fraction of disk galaxies in this class (only 1.9\%), while for the spheroidals they account for 4.6\%, as seen in Tab.~\ref{tab:classes}.
\item \textbf{Class F:} Galaxies of class~F show a similar behaviour to galaxies of class~C, as only half of them experience a major merger and only about 40\% of 
their stellar mass is accreted (see right panel of Fig.~\ref{fig:mass_classes} and upper panel of Fig.~\ref{fig:accretion_hist}). The major difference is the 
amount of gas accreted through the merger events independent of the merger mass ratio: We find that all mergers deliver significantly more gas than for the galaxies of class~C, 
with a total of about 80\% of the stellar mass being accreted in gas mass (see lower panel of Fig.~\ref{fig:accretion_hist}), which is the highest frequency found for the different 
profile classes.
Since all these galaxies have stellar masses well below $10^{11}M_\odot$, they do not host a large hot gas halo, and thus most of that gas is cold and contributes to star formation, resulting in an overall in-situ dominated radial density profile.
This clearly shows that the origin of these overall-in-situ dominated profiles is gas-rich accretion, and as such it is surprising that two of these galaxies are actually spheroidals.
\end{itemize}

\subsection{Accreted Mass Fractions and Galaxy Mass}\label{sec:33}
\begin{figure*}
    \includegraphics[width=.9\textwidth]{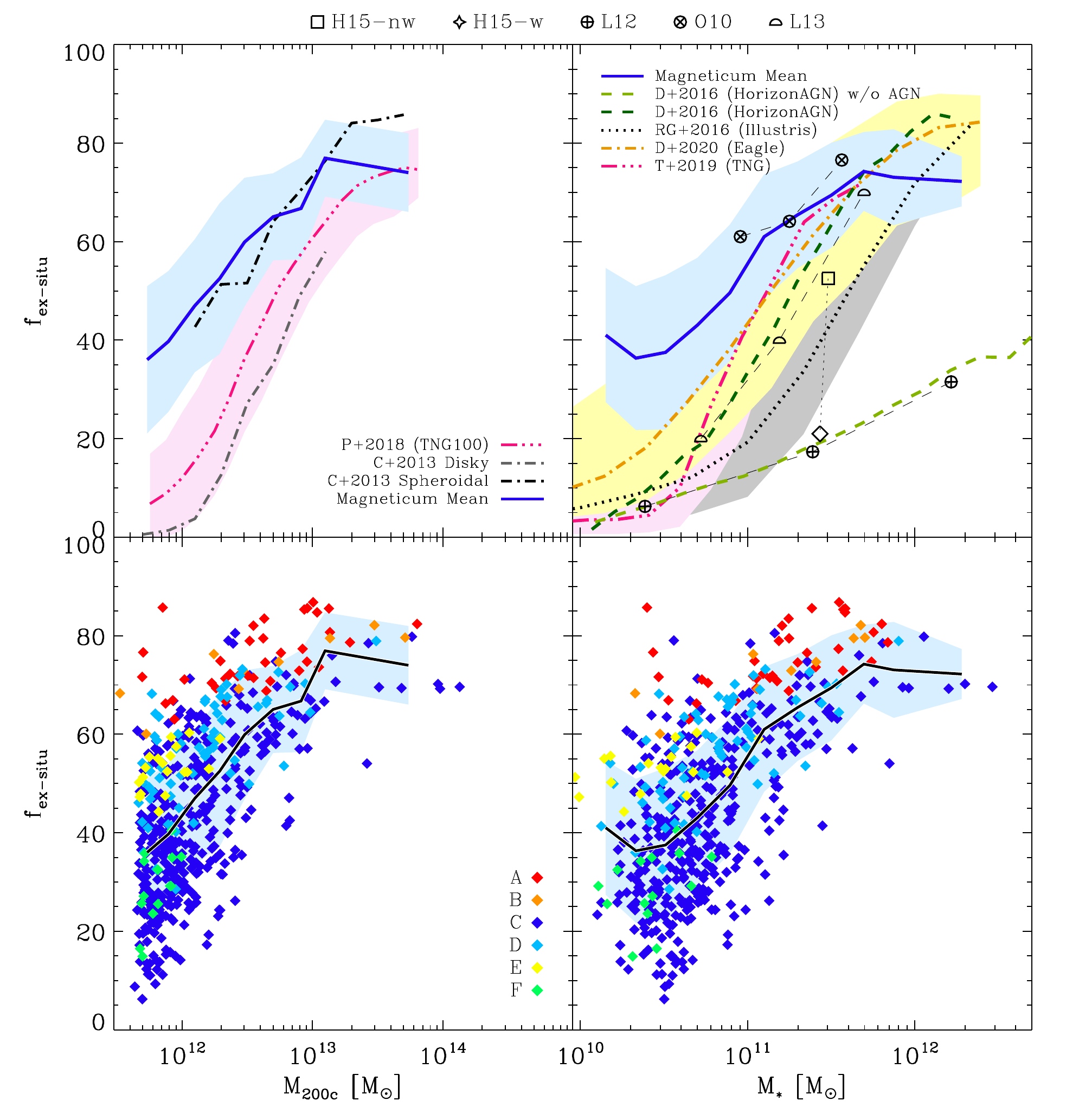}
    \caption{Ex-situ (accreted) fractions for Magneticum galaxies in comparison to other simulations. 
    \textit{Upper left panel:} Mean ex-situ fraction versus critical halo mass $M_\mathrm{200c}$ for Magneticum (solid blue line), with the $1\sigma$ scatter shown in light blue. For comparison, mean values are also shown for Illustris-TNG100 (\citet{pillepich:2018}, dash-dot-dot-dotted pink line and shaded area), and the particle tagging models from \citet{cooper:2013} (dash-dotted lines).
    \textit{Lower left panel:} Same as upper panel but showing the individual values for the Magneticum galaxies, with the colours marking the different profile classes as indicated in the legend. The solid line (here black instead of blue for better visibility) and blue shaded area mark the mean and $1\sigma$ scatter for this distribution, as in the upper panel.
    \textit{Upper right panel:} Ex-situ fraction versus stellar mass $M_{*}$. The blue solid line and shaded area show the mean and the $1\sigma$ scatter for the Magneticum galaxies, as in the left panel. 
    For comparison, we include the relations for four other fully cosmological simulations: Illustris (\citet{rodriguez:2016}, dashed black line and gray shade), Eagle (\citet{davidson:2020}, yellow dash-dotted line and yellow shade), Illustris-TNG (\citet{tacchella:2019}, pink dash-dot-dot-dotted line and shade), and Horizon-AGN (\citet{dubois:2016}, green dashed lines, light green without AGN, dark green with AGN).
    Additionally, we include the mean values from two zoom-simulations that cover a range of stellar masses: the SPH-based GADGET from \citet{oser:2010} as x-circle symbols, and the AMR-based ENZO simulations from \citet{lackner:2012} as +-circle symbols. The mean values in three mass bins from the semi-analytic modeling by \citet{lee:2013} are shown as black half-circles.
    Finally, we also include the mean values for the zoom-simulations by \citet{hirschmann:2015} (square for non-feedback, diamond for the feedback-runs, joined by a vertical line). These data points were extracted from \citet{rodriguez:2016}. We include these simulations as they nicely demonstrate the impact of the stellar feedback on the simulations. 
    \textit{Lower right panel:} Same as the upper right panel, but for the individual Magenticum galaxies with the colours marking the different profile classes as in the lower left panel. Again, the Magneticum mean is shown as solid black line and the blue shaded area marks the $1\sigma$ scatter for this distribution, as in the upper panel. 
    Magneticum galaxies tend to have higher accretion fractions at low masses compared to other simulations.
    }
    \label{fig:fexsitu_comp}
\end{figure*}
Overall, there is broad agreement between simulations and models of different kinds that the fraction of accreted stars is correlated with the stellar (and halo) mass of galaxies.
However, the different simulations vary strongly in the actual (average) values found for the accreted (or ex-situ) fractions at a given mass. 
This can clearly be seen in Fig.~\ref{fig:fexsitu_comp} for halo mass (upper left panel) and for stellar mass (upper right panel). 
Here, we compile data from the literature for the five large hydrodynamical simulations:
\begin{itemize}
\item Magneticum from this study, shown as the blue solid line and shaded area in both panels.
\item Illustris-TNG, shown as the pink dash-dot-dot-dotted line and shaded area, from \citet{pillepich:2018} for the halo mass comparison (left panel) and from \citet{tacchella:2019} for the stellar mass comparison (right panel).
\item EAGLE, shown as dash-dotted yellow line and shaded area, from \citet{davidson:2020}, only for the stellar mass comparison (right panel).
\item Horizon-AGN, shown as green dashed lines, from \citet{dubois:2016}, only for the stellar mass comparison (right panel) for the runs with (dark green) and without (light green) AGN.
\item Illustris, shown as black dotted line and gray shaded area, from \citet{rodriguez:2016}, only for the stellar mass comparison (right panel).
\end{itemize}
As can be seen immediately, Magneticum galaxies have larger accreted fractions at the low mass end compared to the other simulations, while at the high mass end the accreted fractions for all the simulations 
converge to around 70-80\% for galaxies of stellar masses above $3\times 10^{11}M_\odot$ (or halo masses above $1\times 10^{13}M_\odot$).

There are different reasons for these simulations to show such different behaviour, and it is up to now still unclear which ones are closest to reality.
One commonly discussed reason for the different results with respect to the amount of accreted and in-situ components in galaxies is the stellar and/or AGN feedback, which is modeled slightly different in each simulation.
That both processes have strong effects on the resulting accretion fractions has been reported by previous studies: 
\citet{hirschmann:2015} used a subset of the galaxies presented by \citet{oser:2010} and simulated them with and without strong stellar feedback (see open black diamond and square in the upper right panel of Fig.~\ref{fig:fexsitu_comp}, respectively). 
They found that, while the stellar mass of the galaxies did not change much, the accreted fraction dropped significantly, on average from about 50\% to 20\%, when the stellar feedback was switched on.
This is due to the fact that the feedback from the stars heats up the gas inside the galaxies and thus suppresses star formation, especially for the lower mass galaxies, leading to lower stellar masses for the galaxies in general and thus smaller amounts of accreted stars.
However, the merger events still lead to starbursts and subsequent star formation, but this is then in-situ star formation.
An opposite effect was reported by \citet[][see green dashed lines in Fig.~\ref{fig:fexsitu_comp}]{dubois:2016} and \citet{dubois:2013} for the AGN feedback. Here, the simulation runs without AGN feedback (light green) result in lower accreted fractions than the simulation with AGN feedback (dark green).

As all five fully hydrodynamical simulations shown here include both types of feedback, albeit in different implementations, it is not possible to know which of the processes is the main driver of the differences between the simulations.
Note, for example, that the values found for Magneticum are very similar to the average values found by \citet{oser:2010} in their zoom-simulations (x-circles), but the latter has no AGN feedback and the galaxies found in that simulation also differ strongly from those in Magneticum in other parameters \citep[see e.g.,][]{remus:2017}.
For Magneticum, we know from \citet{teklu:2017} that our stellar feedback is slightly too weak for the low mass end and our AGN feedback is slightly too strong at the high mass end when looking at the baryon convergence efficiency,
which is most likely also the reason for the difference at the low mass end between Magneticum and EAGLE and Illustris-TNG.

Another possible reason for the different accretion fractions found in the different simulations could be the use of AMR versus SPH codes.
Those simulations, both fully cosmological and zoom-in, that were performed with AMR codes (namely Horizon-AGN \citep{dubois:2016} and the simulations by \citet{lackner:2012}) show significantly
lower accreted fractions at all mass ranges than those simulations that were performed with SPH codes (namely Magneticum, EAGLE, and the simulations by \citet{oser:2010} and \citet{hirschmann:2014}), independent of the included feedback.
This is especially interesting given that the studies by \citet{lackner:2012} and \citet{oser:2010} both do not include the AGN feedback, but show the strongest differences.
It is well know that all codes have their shortcomings, and many improvements have been implemented in recent years, but to understand how much influence the choice of the code has on the accreted fractions, however, would require a detailed comparison study, which has not been done so far.

Note that, for the five large cosmological simulations shown here, the definitions of in-situ and accreted largely agree in that we only count those stars as accreted that were already born at infall, and count those stars that were born from
gas that was accreted through a merger but only formed after the merger event from this gas as in-situ \citep[see also][]{rodriguez:2016,tacchella:2019}.
This means that, effectively, our accreted fractions are lower limits, and the values could only get higher for more elaborate definitions of in-situ and accreted, but thus this definition
is not responsible for the difference found between these simulation samples.

We also include the accreted fractions with mass obtained from two semi-analytic models (SAMs):
\citet{cooper:2013} used a particle-tagging method on top of the SAM presented by \citet{guo:2011} based on the Millenium II simulation \citep{boylan:2009}, and the resulting accretion fractions for different halo masses
are shown as dash-dotted lines in the left panel of Fig.~\ref{fig:fexsitu_comp}, split according to the morphology as disks (gray line) or spheroidals (black line). The results found for their sample of spheroidal galaxies
is close to the average values found for the Magneticum galaxies in this work, albeit our sample includes both disks and spheroidals. When we separate our disks and spheroidals, we find a general trend for the disks to have lower accreted fractions
than spheroidals of the same mass in agreement with \citet{cooper:2013}, but the trend is much less pronounced and the scatter is large.
\citet{lee:2013} use a SAM built on their own Gadget-2 based dark matter only simulation, and provide accreted fractions for a range of stellar masses (see half-circles in the upper panel of Fig.~\ref{fig:fexsitu_comp}).
Their values lie in between the five different fully cosmological simulations, and are especially at the low-mass end not in agreement with our Magneticum results.

So far, we have discussed the mean values of the accreted fractions with stellar and halo mass for the Magneticum simulation in comparison to other simulations, now we want to take a closer look
at the distribution of the individual galaxies with regard to their accretion classes A--F.
They are shown in the lower two panels of Fig.~\ref{fig:fexsitu_comp} in comparison to the mean value lines shown in black.
As can be seen immediately, there are strong differences between the galaxies of the different accretion classes: 
The galaxies from the over-merged classes A and B all show high accretion fractions, well above 60\%, with no real trend with mass visible. On the other hand, the in-situ dominated galaxies of class~F all have,
as expected, low accretion fractions below the mean Magneticum values, and their spread in mass is too small to see any trend with mass for both stellar and halo mass.
Similarly, galaxies of the major merger class~E also show no trend in mass, and the overall accretion fractions are around 50\%.
For the other two classes, C and D, we find a clear correlation of the accreted fraction with both stellar and halo mass, with a tendency for the 
galaxies of class~D to be slightly above the mean Magneticum accretion values per mass, and for class~C to be slightly below on average.
Interestingly, class~C also includes the lowest accretion fractions at all mass bins, even lower than the galaxies of the in-situ dominated class~F, clearly
demonstrating that the accretion fractions can be really low if most of the accretion is provided by dry minor and mini mergers in the outskirts of a galaxy, while the centre is left undisturbed.
\begin{figure*}
    \includegraphics[width=0.9\textwidth]{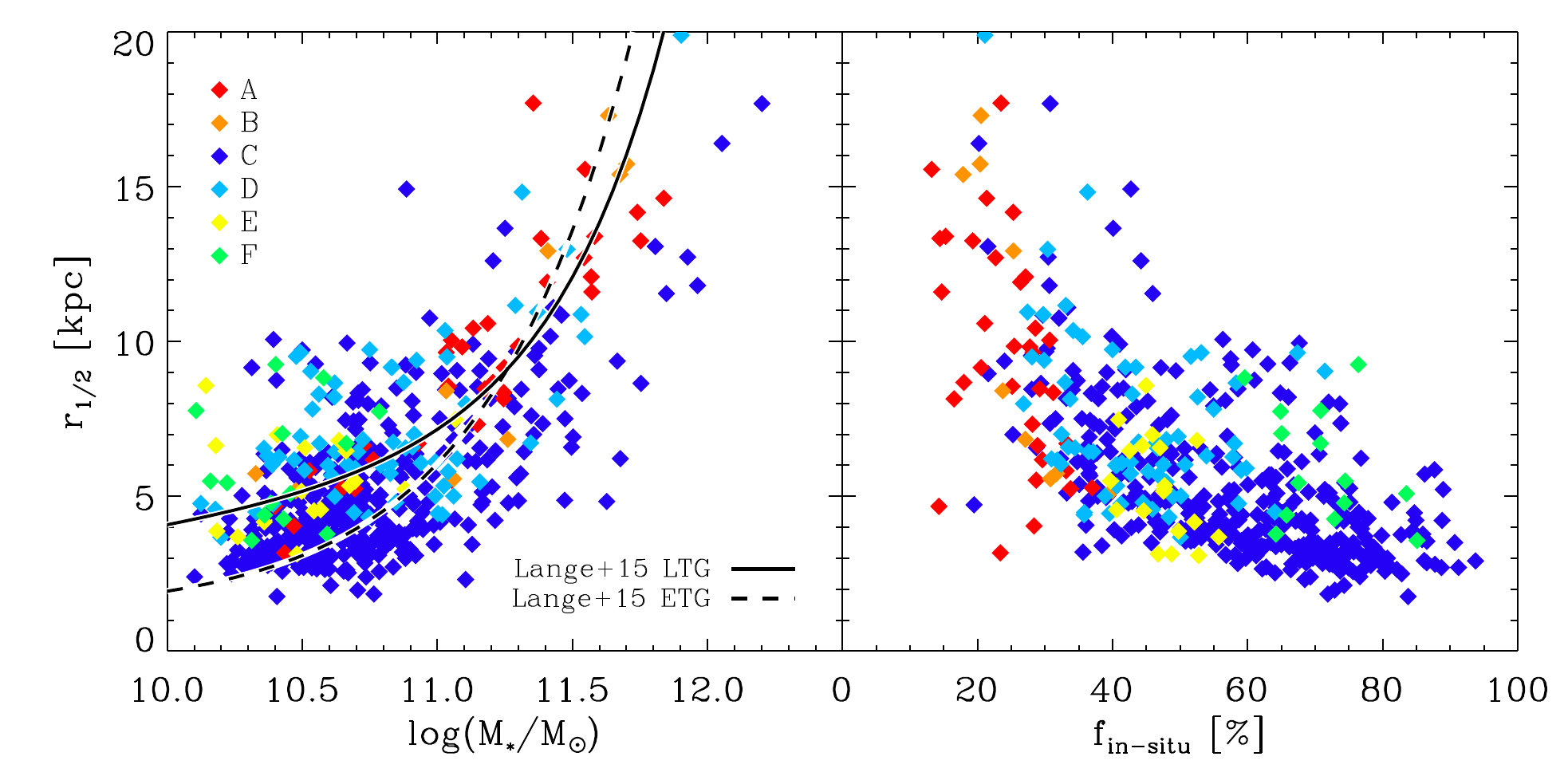}
    \caption{\textit{Left panel:} Stellar mass-size relation for the Magneticum galaxies, colour coded according to their profile class (see legend).
    The 3D half-mass radius is shown against total stellar mass.
    For comparison, the observed
    mass-size relation from the GAMA survey by \citet{lange:2015} is shown for ETGs (dashed line) and LTGs (solid line). 
    Note that the observations measure half-light radii instead of half-mass radii.
    \textit{Right panel:} In-situ fraction versus 3D half-mass radius for the Magneticum galaxies, colour coded as in the left panel.
    Only class~C reveals a clear correlation between in-situ fraction and galaxy size.
        }
    \label{fig:finsitu}
\end{figure*}

\subsection{Accreted Mass Fractions and Global Galaxy Properties}\label{sec:34}
Next we examine trends between the 3D half-mass radius and dark matter fraction with total stellar mass and in-situ fraction. 
It has been shown already by \citet{remus:2017} and \citet{schulze:2018} that the Magneticum galaxies successfully reproduce the observed stellar mass-size relations \citep[e.g., GAMA, by][]{lange:2015}, but here we now
take a closer look at the different profile classes in this relation (left panel of Fig.~\ref{fig:finsitu}).
The overmerged galaxies of classes A and B show only small scatter close to the observed relation over the whole mass range, but due to the fact that they are
the by far most common class at high stellar masses, they are also most common among the large galaxies.
The classical profile galaxies of class~C, however, show a significantly different behaviour from the galaxies of the other classes, in that they are the clearly
dominant class for the small galaxies at all stellar mass ranges. Albeit the scatter is large for this class and there are also very large galaxies among them,
they clearly dominate the small size end especially at the lower stellar mass end. This reflects our previous findings that these galaxies are dominated by compact star formation
in their centers and only little accretion mostly to the outskirts, resulting in a rather compact central part and consequently a smaller half-mass radius.
On the other hand, we see a very different behaviour for those galaxies of classes D, E, and F, all of which have large sizes for their stellar masses, clearly dominating the
region of the mass-size relation that is usually occupied by disk galaxies. This is in good agreement with the fact that all of them have large amount of cold gas accreted through their 
formation history since z=2, resulting in in-situ star formation in disks and thus larger half-mass radii (even if in case of class~D galaxies the large central accreted component will
prevent its classification as a disk given its massive bulge-like nature).

As stellar mass $M_*$ and 3D half-mass radius $r_{1/2}$ of a galaxy are correlated, it is not surprising that we also find a correlation between the in-situ fraction of a galaxy and its half-mass radius (right panel of Fig.~\ref{fig:finsitu}).
It can best be seen for the galaxies of accretion class~C, as they cover the largest range of both half-mass radii and in-situ fractions, with a clear tendency for smaller galaxies to have larger in-situ fractions and large galaxies to have small in-situ fractions. A similar behaviour is found for galaxies of classes D and E, albeit class~E only covers such a small range of in-situ fractions that no clear correlation between size and in-situ fraction can be inferred from these galaxies alone.
In general, the in-situ fraction decreases with increasing halfmass radius, i.e., accretion leads to a growth in the scaled size \citep[e.g.,][]{oser:2012}, which indicates that most of the accreted stars are deposited at large radii \citep[see also][]{amorisco:2017,lagos:2018,karademir:2019,davidson:2020}.

\begin{figure*}
    \includegraphics[width=0.95\textwidth]{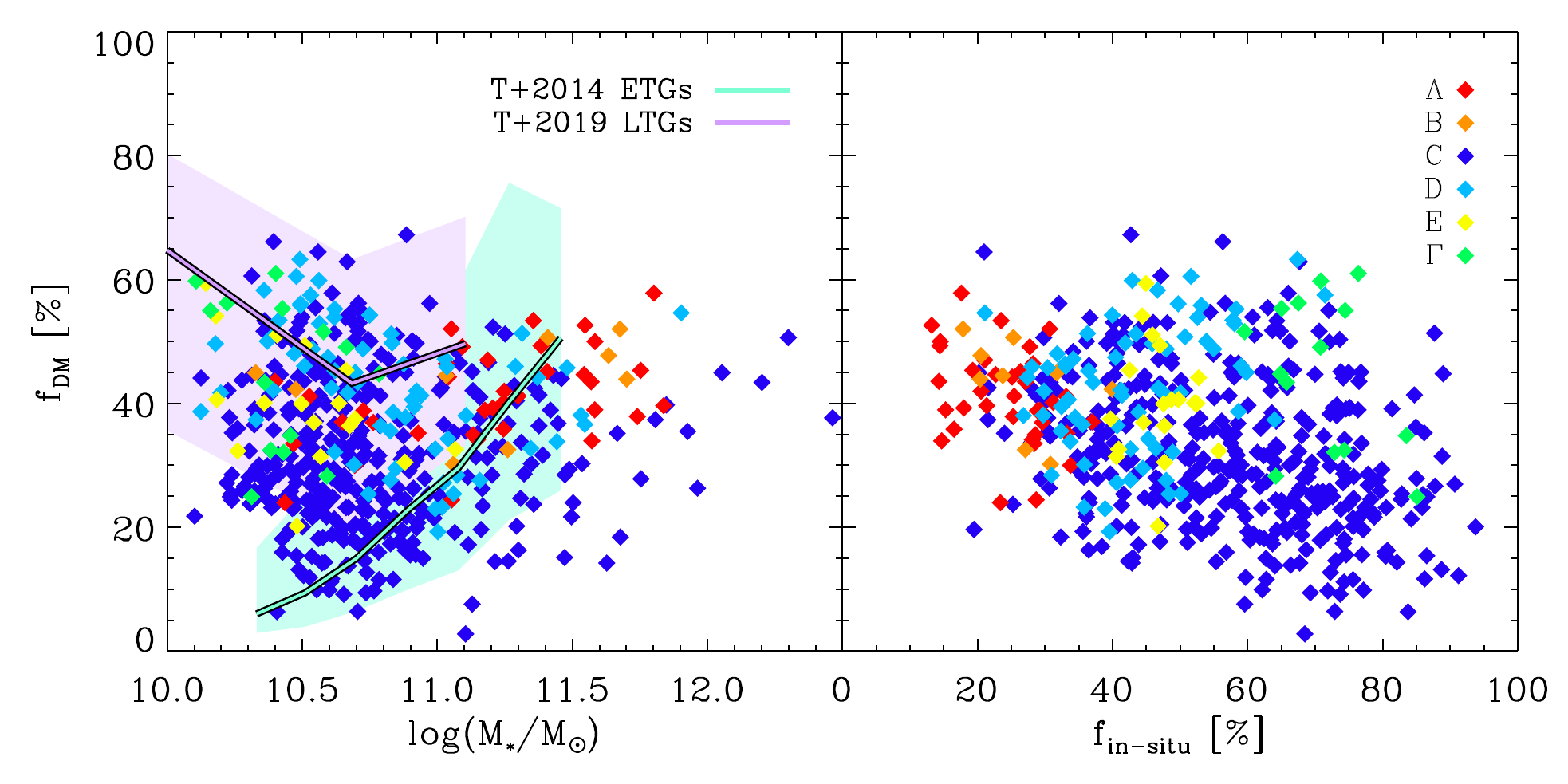}
    \caption{Dark matter fraction within the half-mass radius trends.
    \textit{Left panel:} Dark matter fraction $f_\mathrm{DM}$ versus stellar mass $M_*$, with colours as in the right panel.
    Observations for LTGs from the SPARCS survey \citep{tortora:2019} are included as a lilac solid line and shaded area, and observations for ETGs from the SPIDER survey \citep{tortora:2014} are included as an aqua solid line and shaded area.
    \textit{Right panel:} Dark matter fraction $f_\mathrm{DM}$ versus in-situ fraction $f_\mathrm{in-situ}$ for the Magneticum galaxies, with colours indicating the different accretion classes.
    }
    \label{fig:fdm}
\end{figure*}
For galaxies of the overmerged classes A and B we find a similarly large range of half-mass radii, but only a small range of in-situ fractions around 20\%, and they reveal no correlation at all between size and in-situ fraction. This well reflects
the known fact that a major merger results in a much more compact galaxy than a series of minor mergers that bring in the same total mass as the major merger but deposit their masses at different radii \citep{naab:2009,hilz:2012}. So while all 
galaxies of these two classes had plenty of mergers, we find the differences in the individual merger mass ratios mirrored in the size distribution.
Galaxies of the in-situ dominated class~F show the strongest deviation from the correlation between size and in-situ fraction: while all the in-situ fractions are rather high, the sizes are generally
larger than those of the class~C galaxies of similar in-situ fraction, in agreement with our previous finding that class~F galaxies are more similar to disks than the average class~C galaxy.

Finally, we investigate if the fraction of dark matter within the half-mass radius, $f_\mathrm{DM}$, is correlated with the in-situ fraction and stellar mass.
As can be seen in Fig.~\ref{fig:fdm}, there is a broad tendency for galaxies with smaller in-situ fractions to have larger central dark matter fractions, indicating that
(massive) accretion events lead to larger fractions of dark matter in the center by either enhancing the relative amount of dark matter in the center or dispersing the baryonic matter.
This tendency can be seen for galaxies of all accretion classes but those of class~F, the in-situ dominated class. Galaxies of that
class show much higher central dark matter fractions than galaxies of class~C with similar in-situ fractions.
This is in good agreement with our previous conclusion that class~F galaxies closely resemble the typical behaviour of disk galaxies,
since observations show that LTGs have, at the same stellar mass, larger central dark matter fractions than ETGs \citep[][but also e.g., \citet{courteau:2015,genzel:2020}, albeit they use velocity dispersions instead of stellar masses and different definitions of central radius]{tortora:2019}.
This can also be seen in the left panel of Fig.~\ref{fig:fdm} where we included the observational results for LTGs and ETGs from \citet{tortora:2019} and \citet{tortora:2014}, respectively.
As can immediately be seen, most of the class~D and F galaxies clearly resemble the properties of the LTGs, while the clear observed correlation between $f_\mathrm{DM}$ and $M_*$ for ETGs is most strongly populated by galaxies of the classical accretion profile class~C, in good agreement with the idea that dry merging lowers the central dark matter fractions while wet merging and smooth gas accretion lead to larger central dark matter fractions.
However, the details of the interactions between the baryons and the dark matter in the centers of galaxies and the influence of gas and feedback on this interaction are currently under debate and are beyond the scope of this work.
 
\subsection{Accreted Mass Fractions and Transition Radii}\label{sec:35}
\begin{figure*}
	\includegraphics[width=0.9\textwidth]{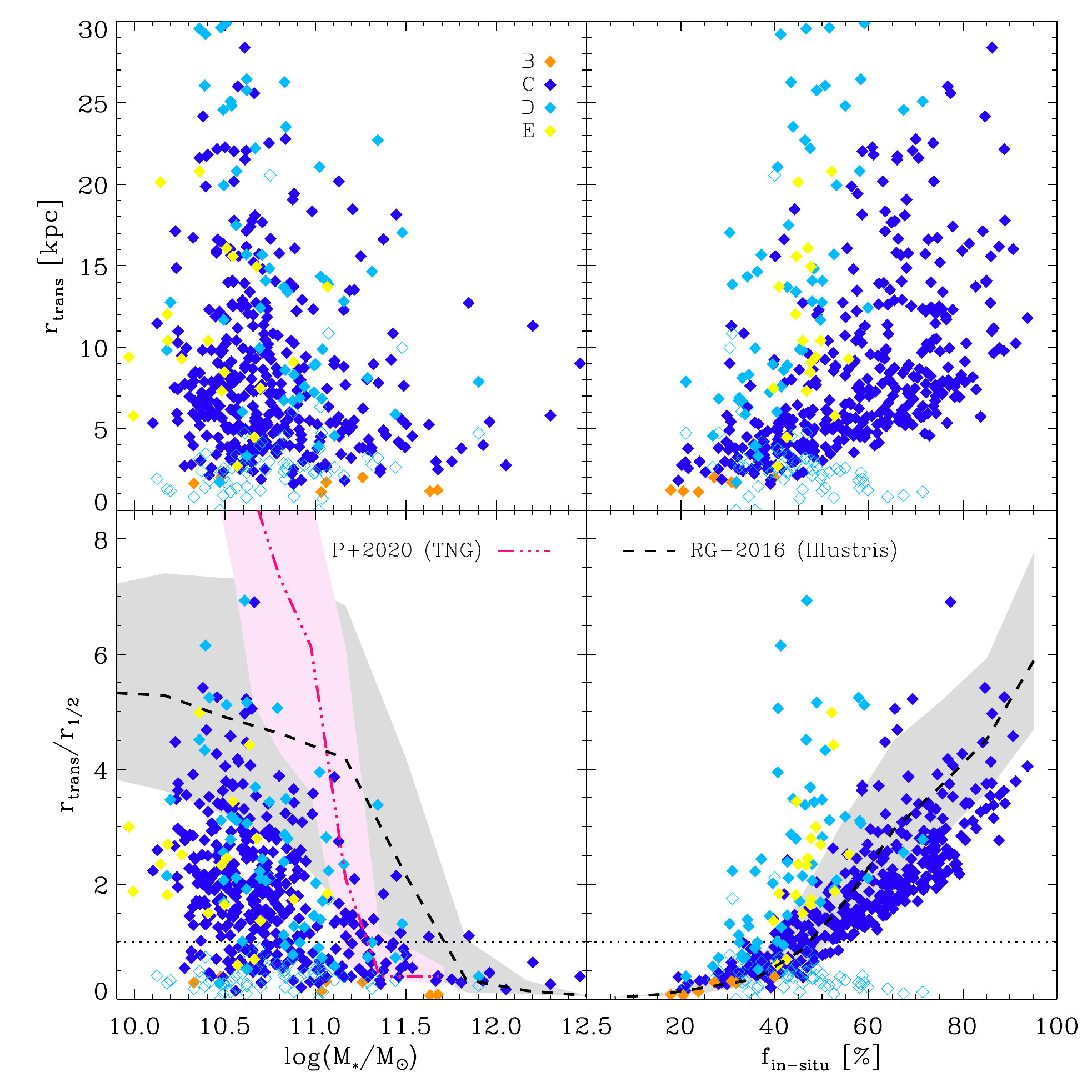}
    \caption{Transition radius trends for all profile classes as indicated in the label. Classes A and F have no transition radii and are therefore not shown.
    Class~D galaxies (cyan diamonds) have two transition radii, so the outer ones are shown as filled diamonds and the inner ones are shown as open diamonds.
    \textit{Upper left panel:} Stellar mass $M_*$ versus 3D transition radius $r_\mathrm{trans}$ in kpc.
    \textit{Upper right panel:} In-situ fraction $f_\mathrm{in-situ}$ versus transition radius $r_\mathrm{trans}$ in kpc.
    \textit{Lower left panel:} Stellar mass $M_*$ versus 3D transition radius $r_\mathrm{trans}$ normalised by 3D half-mass radius $r_\mathrm{1/2}$.
    \textit{Lower right panel:} In-situ fraction $f_\mathrm{in-situ}$ versus normalised transition radius. 
    The horizontal line in both lower panels represents a normalised transition radius of 1 half-mass radius, i.e., $r_\mathrm{trans}/r_\mathrm{1/2} = 1$.
    Dashed black lines and shaded areas show the results from the Illustris simulation \citep{rodriguez:2016}, and the dash-dot-dot-dotted 
    pink line and shaded area are the results found for Illustris-TNG \citep{pulsoni:2020}.
    The horizontal dotted line marks where the transition radius equals the half-mass radius. 
	}
    \label{fig:rtrans}
\end{figure*}
As discussed before, for most galaxies there exists a radius at which the contribution from in-situ and accreted stars is 50\% each, that is at which the dominance of the two components switches.
We call this radius the \textit{transition radius} $r_\mathrm{trans}$.
For our classic profile (class~C), this is the radius where the dominant stellar component switches from in-situ in the center to accreted in the outskirts, and thus separates the inner, self-made part of the galaxy from the outer, dry-merger dominated part.

Previous works by \citet{cooper:2013} and \citet{rodriguez:2016} already reported this radius to be 
smaller for larger stellar masses and smaller in-situ fractions, and we can confirm these general trends for our class~C galaxies as shown in Fig.~\ref{fig:rtrans}.
However, we do not find a tight correlation between the transition radius and stellar mass, and only a weak correlation is seen between transition radius and in-situ fraction (see upper panels of Fig.~\ref{fig:rtrans}), with a large scatter.
Only when moving to normalised transition radius (i.e., the transition radius divided by the half-mass radius), the trends become more clear: we even see a clear positive correlation between
normalised transition radius and in-situ fraction, very similar to the correlation found by \citet{rodriguez:2016} but slightly less steep (see lower panels of Fig.~\ref{fig:rtrans}). 
We also find a clear negative trend between the in-situ fractions and the stellar mass, with galaxies that have accreted a lot of material (i.e., high mass galaxies) tending to have normalised 
transition radii of $r_\mathrm{trans}/r_\mathrm{1/2} \le 1$. However, this trend is more an upper limit for the in-situ fractions at a given mass, as we also find low-mass galaxies
with normalised transition radii $r_\mathrm{trans}/r_\mathrm{1/2} \le 1$, but basically no high-mass galaxies with $r_\mathrm{trans}/r_\mathrm{1/2} > 1$.
The trend found for the Magneticum galaxies is weaker than what has been found by \citet{rodriguez:2016}, and much weaker than the trend reported for the Illustris-TNG galaxies by \citet{pulsoni:2020}.
This reflects the result found already in Fig.~\ref{fig:fexsitu_comp}, namely that the galaxies in Magneticum have, on average, accreted more dry stellar mass through mergers than galaxies in galaxies from Illustris or Illustris-TNG.

So far, we only discussed those galaxies of profile class~C as most previous works only discussed this profile class with no mention of other profile classes.
For classes A and F, we cannot provide a transition radius as these galaxies are always dominated by accreted or in-situ stars, respectively, but for the profile classes B, D, and E such transition radii exist: 
Galaxies of class~B usually have a very small transition radius of only about $2~\mathrm{kpc}$, close to the limits of our spatial resolution (and hence may be somewhat smaller than indicated). 
We do not find any trend, positive or negative, with stellar mass, and only a weak positive correlation between in-situ fraction and normalised transition radius.
This is not surprising as this class if very close to being overmerged like class~A, and thus we do not expect the transition radius to have any relevant meaning.

In the case of class~D, where the center and the outskirts are dominated by accreted stars but the middle radial range is dominated by in-situ stars, even two transition radii exist.
For the outer transition radii of class~D (filled cyan symbols in Fig.~\ref{fig:rtrans}) and class~E we find the trend with in-situ fraction to be very similar, but generally steeper than the correlation seen for class~C. 
This is even clearer for the normalized transition radii again. For both classes we also find a tighter anti-correlation between normalised transition radius and stellar mass, albeit the scatter is still large.

The inner transition radii for class~D galaxies (open cyan diamonds in Fig.~\ref{fig:rtrans}) are usually comparably small and often well below the half-mass radius. 
Generally, the inner transition radii of class~D behave significantly different from all other transition radii, as there is no trend at all for the stellar mass neither with the 
transition radius nor with the normalised transition radius, and there is actually a negative trend with in-situ fraction, clearly showing that the larger the fraction of stars formed in-situ, the smaller the accreted core in the centre, indicating that more stars are formed in-situ if the mass accreted onto the centre was small compared to the gas disk of the progenitor galaxy.

As these transition radii are very indicative of the accretion history of the galaxies and may provide a method to estimate the in-situ fraction of a galaxy, it would be
very instructive to be able to measure this transition radius observationally. Therefore, in the next section of this paper we address the question of whether it is possible to measure the radius of the transition from in-situ to accretion dominance from the observed surface brightness profiles of galaxies, as suggested by \citet{cooper:2013} and \citet{rodriguez:2016}.

\section{Accreted Fractions from S\'ersic Fits for ETGs}\label{sec:4}
\begin{figure*}
    \includegraphics[width=0.28\textwidth]{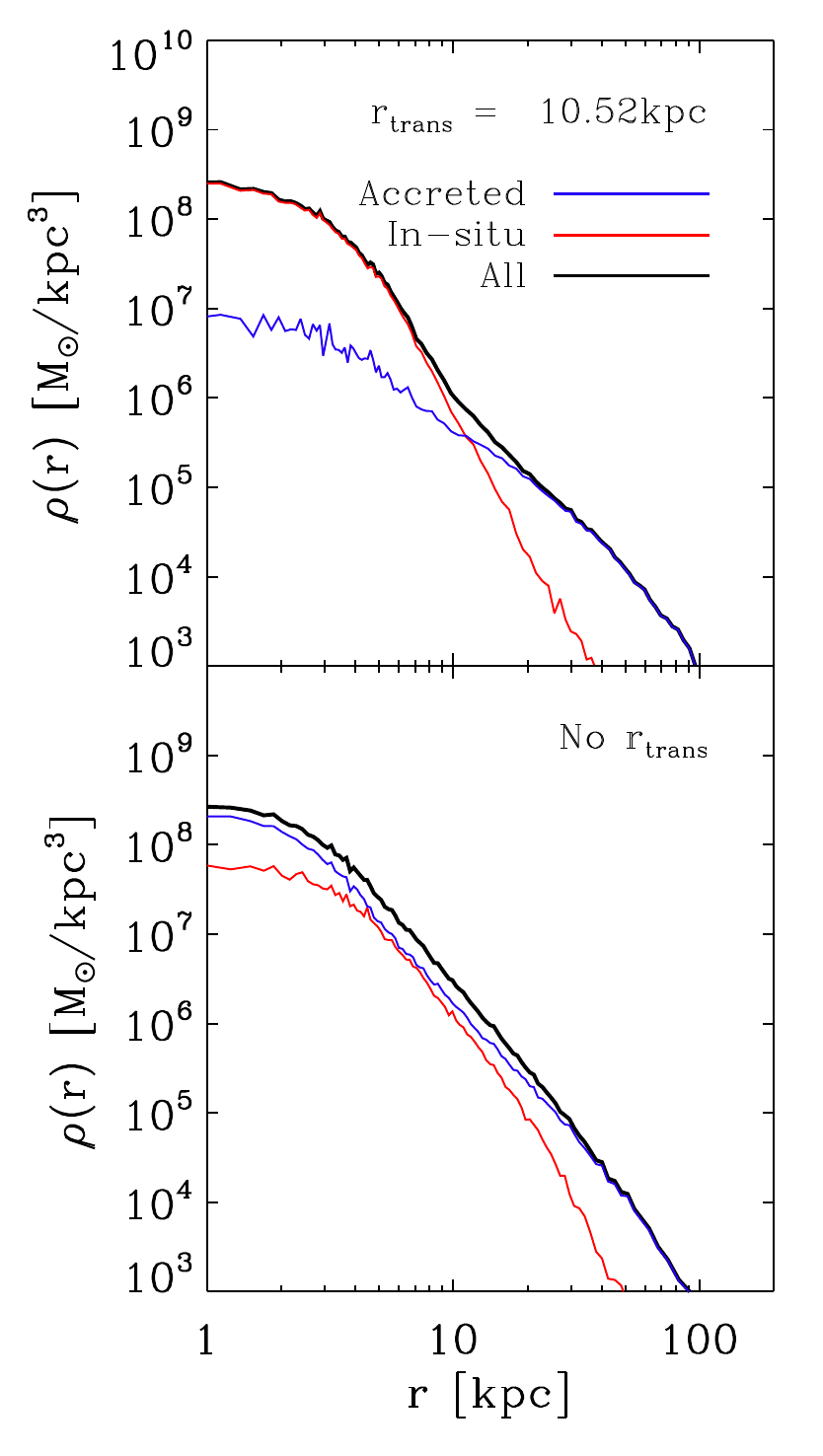}
    \includegraphics[width=0.70\textwidth]{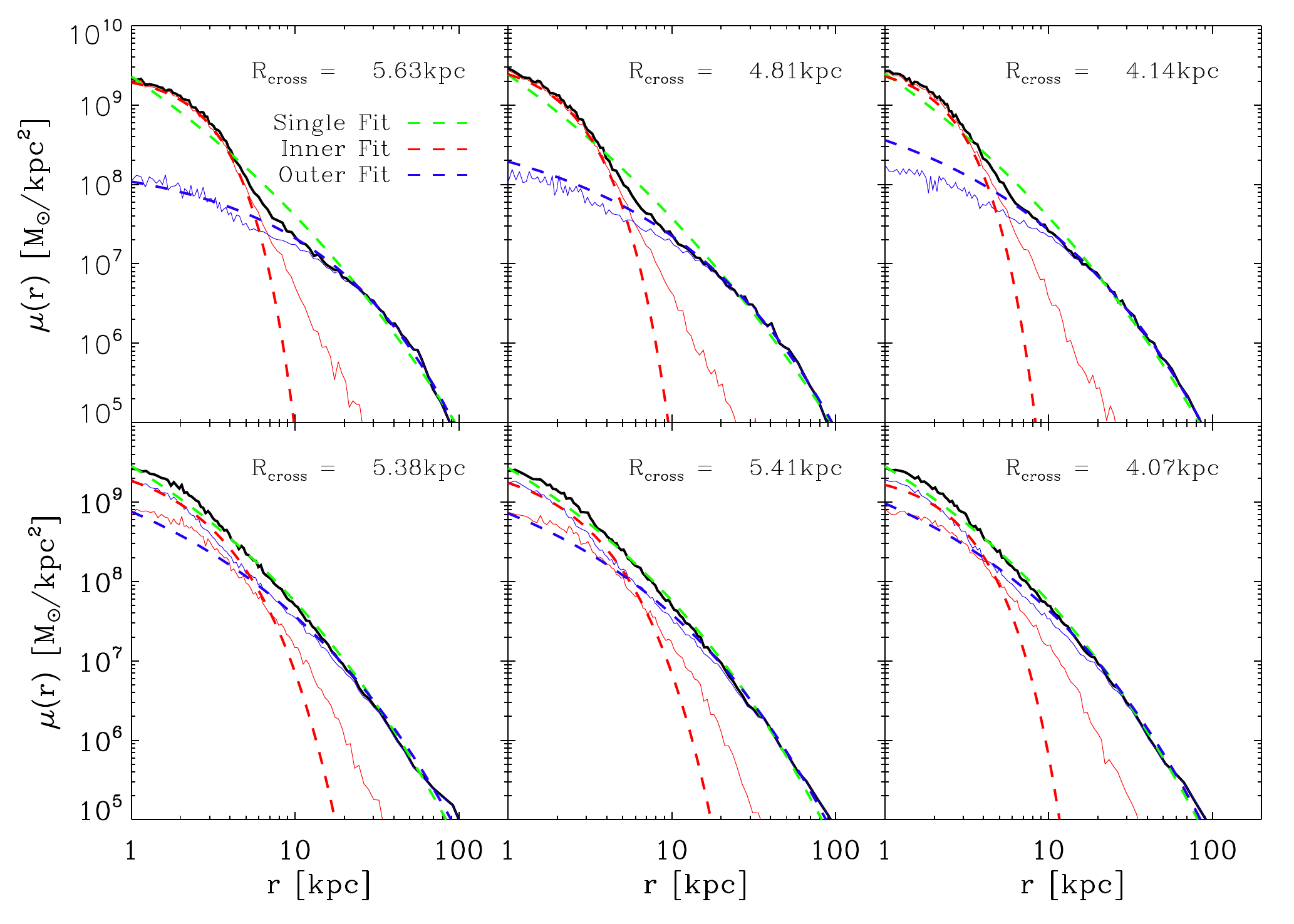}
    \caption{\textit{Upper panels:} Example of a class~C mass density profile in 3D and 2D projections.
        \textit{Top left panel:} 3D total stellar mass density profile (black curve) with the in-situ and accreted components in red and blue, respectively.
        \textit{Top right panels:} Projected 2D stellar mass density profiles from three different projections: The projected total stellar profile is shown as black curve, the in-situ and accreted components are shown as solid red and blue lines, respectively. 
	The dashed lines show the inner (red) and the outer (blue) fits from the double S\'ersic fits, and the single S\'ersic fits (green) to the total projected stellar density profiles. 
	In the upper right of each panel we list the transition radius in 3D and the crossing radii in 2D, in units of kpc.  
    In this example, the single S\'ersic fit is never a good fit to any of the projections. The double S\'ersic fits describe the total profile very well in all projections, and are also a good approximation to the in-situ and accreted profiles in all cases.
    However, the crossing radii $R_\mathrm{cross}$ (i.e., the radius where the two S\'ersic profiles cross) vary on the order of 1~kpc
	between the three projections, and are in all three cases only about half as large as the real 3D transition radius $r_\mathrm{trans}$.
	\textit{Lower panels:}
	Same as upper panels but for a class~A profile. Class~A galaxies are extremely accretion dominated and have no transition radius between in-situ and accreted components in their 3D or 2D stellar density profiles, as can be seen from the solid red and blue curves in all four panels. However, a single S\'ersic fit is not a good fit to any of the projections and the double S\'ersic fit is clearly needed in all three projections to describe the total total stellar profiles. Thus, we obtain crossing radii $R_\mathrm{cross}$ from these double S\'ersic components, that vary again between the three projections, but in no case are they representative of the true in-situ or accreted components.
	}
    \label{fig:example_proj_c}
\end{figure*}

Motivated by previous simulation results from \citet{cooper:2013} and \citet{rodriguez:2016}, there have been observational attempts to measure the in-situ and accreted fractions of galaxies using a double
S\'ersic fit \citep{sersic:1963} to the observed surface brightness profiles of galaxies.
This assumes that the inner S\'ersic fit describes the in-situ component of the galaxy, and the outer S\'ersic fit describes the accreted component.
In some cases, a third fit to the very outskirts of a galaxy was carried out, under the assumption that the third component describes the stellar halo of the galaxy and not the galaxy itself \citep[e.g.,][]{spavone:2017}, but we will not investigate this approach here.
Instead, we investigate in this section from our simulated galaxies if the double-S\'ersic approach really supplies a good measure for the in-situ and accreted components of galaxies.
As the observational work has so far focused mostly on early-type galaxies (ETGs), we restrict our sample to Magneticum ETGs only.
\begin{figure*}
    \includegraphics[width=0.9\textwidth]{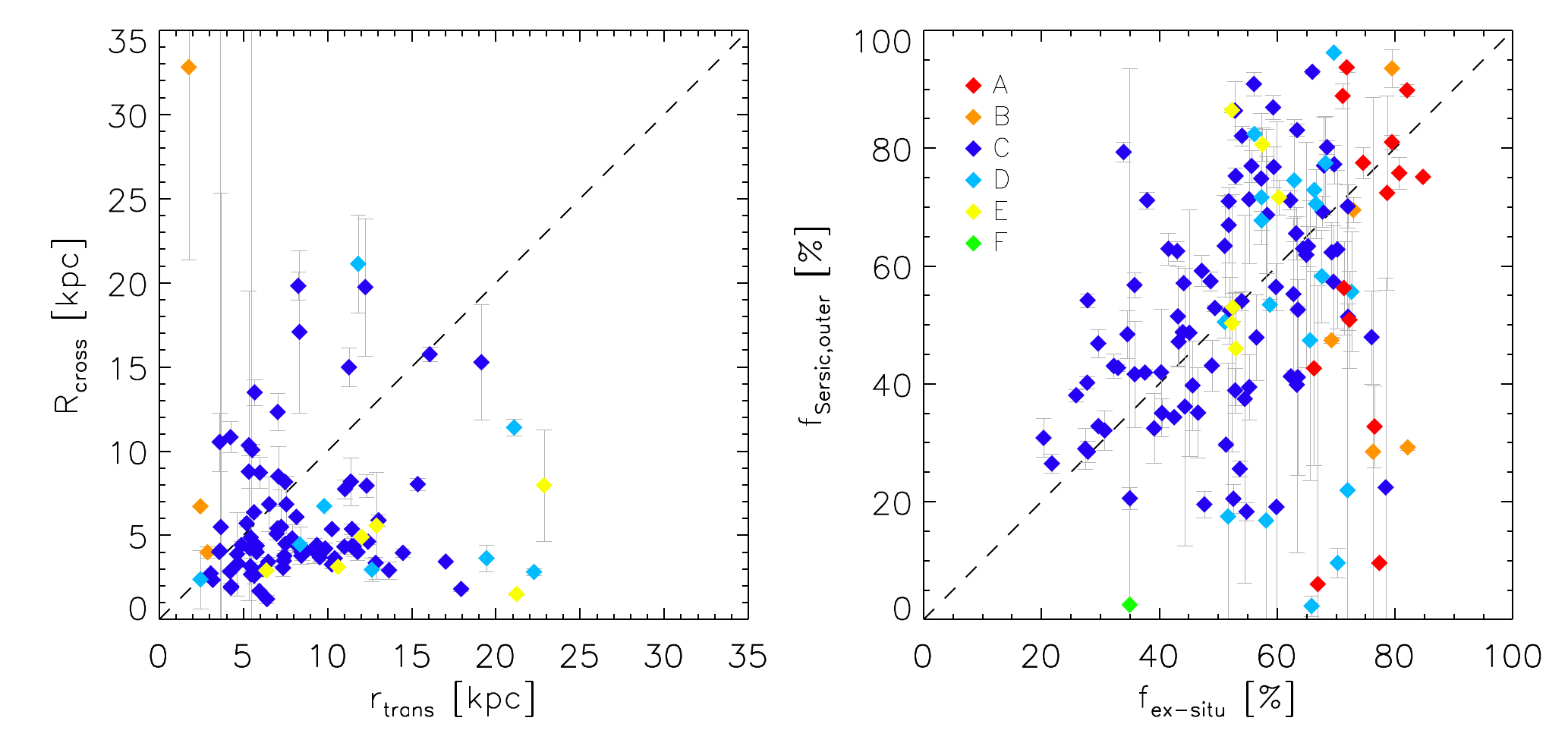}
    \caption{\textit{Left panel:} Crossing radius $R_\mathrm{cross}$ obtained from  double S\'ersic fits to the 2D projected mass density profiles (random projection) versus the 3D transition radius $r_\mathrm{trans}$ between in-situ and accreted components, for all profile classes (colours as indicated in the right panel) with well defined transition radii.
    \textit{Right panel:} Fraction of integrated mass from the outer S\'ersic function fit to the 2D mass density profiles versus the true accreted mass fraction.
    In both panels, the dashed line shows a 1:1 relation. Error bars indicate the maximum and minimum values obtained for edge-on and face-on projections.
        }
    \label{fig:rcross}
\end{figure*}

\subsection{S\'ersic Fits to Projected Surface Density Profiles of Simulated ETGs}\label{sec:41}
To compare the simulated mass density profiles with observed surface
brightness profiles, we need to create 2D projections of the simulated
galaxies. As this projection is rather arbitrary, we choose a random projection along with the face-on and edge-on projections to test for each galaxy in our sample. 
In all cases, we find that the profile class of our galaxies does not change, and the in-situ to accreted relations stay the same under all projections.
Thus, for all galaxies that have a transition radius ($r_\mathrm{trans}$) in 3D, we also find a radius in the 2D projections which indicates a transition from in-situ to accretion dominance.

To follow the observational approach, we fit the projected surface density profiles with both single and double S\'ersic fits.
In most cases, a double S\'ersic fit is a better fit to the projected surface density profiles, independent of the projection.
In those cases where a single S\'ersic fit is sufficient, this is true for all tested projections.
This is rather promising as this clearly indicates that, if a double S\'ersic fit is needed to describe the observed surface brightness profile, then it is independent of the viewing angle and reflects
the underlying 3D density distribution.
For the double S\'ersic fits to the 2D surface density profiles, we define the crossing radius, $R_\mathrm{cross}$, as the radius where the inner and outer S\'ersic profiles cross each others.

The upper row of Fig.~\ref{fig:example_proj_c} shows an example of a class~C profile galaxy with its well defined transition radius in 3D (here $r_\mathrm{trans} = 10.52~\mathrm{kpc}$, left panel).
A transition is also seen between the in-situ and accretion dominated regions of the galaxy in all three 2D projections (upper right panels), however, the values of the transition radii in the 2D projections are all smaller than the 3D transition radius $r_\mathrm{trans}$.
We find that this is not simply a matter of unlucky projections but is rather a common feature of class~C profiles (which make up the majority of profiles).
This disconnect between the transition radii seen in 3D and the 2D profiles also occurs in all other classes with well-defined transition radii, namely class B, D, and E.

While the transition radii are already disconnected from 3D to 2D, the matter is even worse if we use the double S\'ersic fits to describe the underlying in-situ and accreted components: In a few cases like the one shown in the upper panels of Fig.~\ref{fig:example_proj_c}, the two S\'ersic components are a good approximation of the in-situ and accreted components, and the resulting crossing radius between the two S\'ersic components, $R_\mathrm{cross}$, is a good approximation to the 2D transition radius.
However, for most galaxies this is \textit{not} the case.
One example of a galaxy that demonstrates the issue nicely is shown in the lower panels of Fig.~\ref{fig:example_proj_c}:
This galaxy is of class~A, i.e., is accretion dominated at all radii and has no transition radius from in-situ to accretion dominated, neither in 3D (left lower panel) nor in projection (three panels on the lower right).
However, the stellar 2D surface density profiles in this example, under all projections, clearly require a double S\'ersic fit, thus providing a crossing radius $R_\mathrm{cross}$.
The two resulting S\'ersic components in this case do \textit{not} describe the underlying in-situ and accreted components. They instead mark the radius where accretion due to massive mergers transitions into accretion from small mergers and mini mergers that never reach the centre of the galaxy.

To further quantify this issue, the left panel of Fig.~\ref{fig:rcross} shows the differences between the crossing radii of the double S\'ersic fits $R_\mathrm{cross}$ for the random projection (with error bars marking the values for the edge-on and face-on projections), and the true 3D transition radius $r_\mathrm{trans}$ between the in-situ and accreted components for all galaxies where
such a transition radius is well defined (classes B, C, D, and E).
The plot is largely a scatter diagram with little, or no, correspondence between the two measured radii, independent of the profile class.
This is also true for the projected 2D transition radii, but as the behaviour is nearly identical we do not show this plot here.

For galaxies of classes~D and E, the 2D crossing radii are much smaller than the real transition radii, while for galaxies of class~B we find the opposite trend (with two of them having such large crossing radii
$R_\mathrm{cross}$ that they are well above the plotted radius range).
Galaxies of class~C show both kind of behaviour, with $R_\mathrm{cross}$ both lower and higher than the true $r_\mathrm{trans}$.
Independent of the profile class, we conclude that it is a lucky coincidence if the crossing radius $R_\mathrm{cross}$ of the double S\'ersic fits is a good approximation to the transition radius
$r_\mathrm{trans}$.
In summary, the transition from an inner to an outer S\'ersic fit to an observed surface brightness profile bears little, or no, connection to the true transition between the in-situ and accreted components
in a galaxy.

We also measure the integrated mass within the outer S\'ersic component and compare it to the true accreted mass fraction.
This is shown in the right panel of Fig.~\ref{fig:rcross} for all galaxies where a double S\'ersic fit was
a good fit, even those of profile classes A and E that are dominated by accreted or in-situ stars at all radii, respectively.
This plot reveals a large scatter with a very weak trend about the unity line, indicating that an outer (inner) S\'ersic component fit to a surface brightness profile is a poor guide to the true accreted (in-situ) mass
fraction.

In summary, we find that fitting a double S\'ersic profile to the 2D surface density profile of an ETG will \textit{not} reveal the true radius for the transition from in-situ to accretion dominated material.
This suggests that the dips seen in observed surface brightness profiles can not, in general, be taken as a signature of a division between in-situ and accreted components of the galaxy.
They may instead be more indicative of a transition from stars being formed in-situ plus stars accreted by major mergers, to the component of stars mostly accreted through minor or mini mergers.
Furthermore, the integrated mass associated with the inner and outer S\'ersic functions provide a poor guide to the true in-situ and accreted mass fractions of a galaxy, respectively.
We conclude that the (two) components visible in the (projected) density profiles do not reflect the in-situ and accreted components in general, but rather mark the transition from the inner part of the
galaxy which can be dominated by in-situ stars but also be dominated by a massive merger event, and the outer part of the galaxy which is dominated by small minor or mini mergers that get disrupted in the
outskirts of the galaxy and never interact with its center\footnote{Note that in the very inner parts of galaxies additional components can be visible in the (projected) density profiles, caused for example by bars and bulges, but we cannot include these structures in our analysis as the resolution of the Magneticum galaxies is not high enough to resolve these inner structures of the galaxies.}.
This is similar to the dynamical split of the ICL and the BCG in galaxy clusters, and might be a way to distinguish outer stellar halos of galaxies from the galaxies themselves instead.

\subsection{Observational Comparison Samples}\label{sec:42}
Some of the deepest imaging of nearby ETGs available comes from the VEGAS survey \citep{capaccioli:2015}.
The survey probes surface brightness profiles out to $\sim 10 R_\mathrm{e}$ and down to surface brightness levels of $\sim 29~\mathrm{mag/arcsec}^2$ in the g-band.
The survey is still ongoing, however, results on the radial surface brightness profiles have been published for several massive galaxies in group/cluster environments by \citet{spavone:2017} and \citet{spavone:2020}.
\citet{spavone:2017} fit two or three S\'ersic profiles to 6 ETGs, with the S\'ersic parameters $n$ constrained to a narrow range. More recently, \citet{spavone:2020} fit 19 ETGs in the Fornax cluster
with either two or three S\'ersic components. 
Here we focus on the two component fit, for which $n$ was a free parameter.
The two component fits have a single intermediate radius, and the accreted mass fractions are calculated from the second (outer) component.
These are referred to as the ``relaxed'' components following \citet{cooper:2015BCG}.
Hence the approach to provide unconstrained fits is more comparable to our 
approach, we focus on the study by \citet{spavone:2020} instead of \citet{spavone:2017}.

Another very deep imaging study has been carried out by \citet{kluge:2019}, who fit double S\'ersic profiles to extremely low surface brightness profiles targeting especially BCGs. Both single and double S\'ersic fits were obtained, as well as accreted fractions from the double S\'ersic fits. 

We also compare to the double S\'ersic fits of $\sim 45,500$ galaxies, observed at a mean redshift $z \sim 0.08$ and stacked in mass bins by \citet{dsouza:2014}.
Taking mean values from their figure~13 for ETG-like galaxies, we note that their data covers a similar stellar mass range to our modelled galaxies and that they find effective radii of the inner and outer
components to be around 3 and 8~kpc, respectively.
\citet{dsouza:2014} also provided such measurements for their stacked LTG sample, however, we will focuss on their results regarding the ETGs here.

\subsection{Accreted Fractions from Double S\'ersic Fits}
\begin{figure}
    \includegraphics[width=\columnwidth]{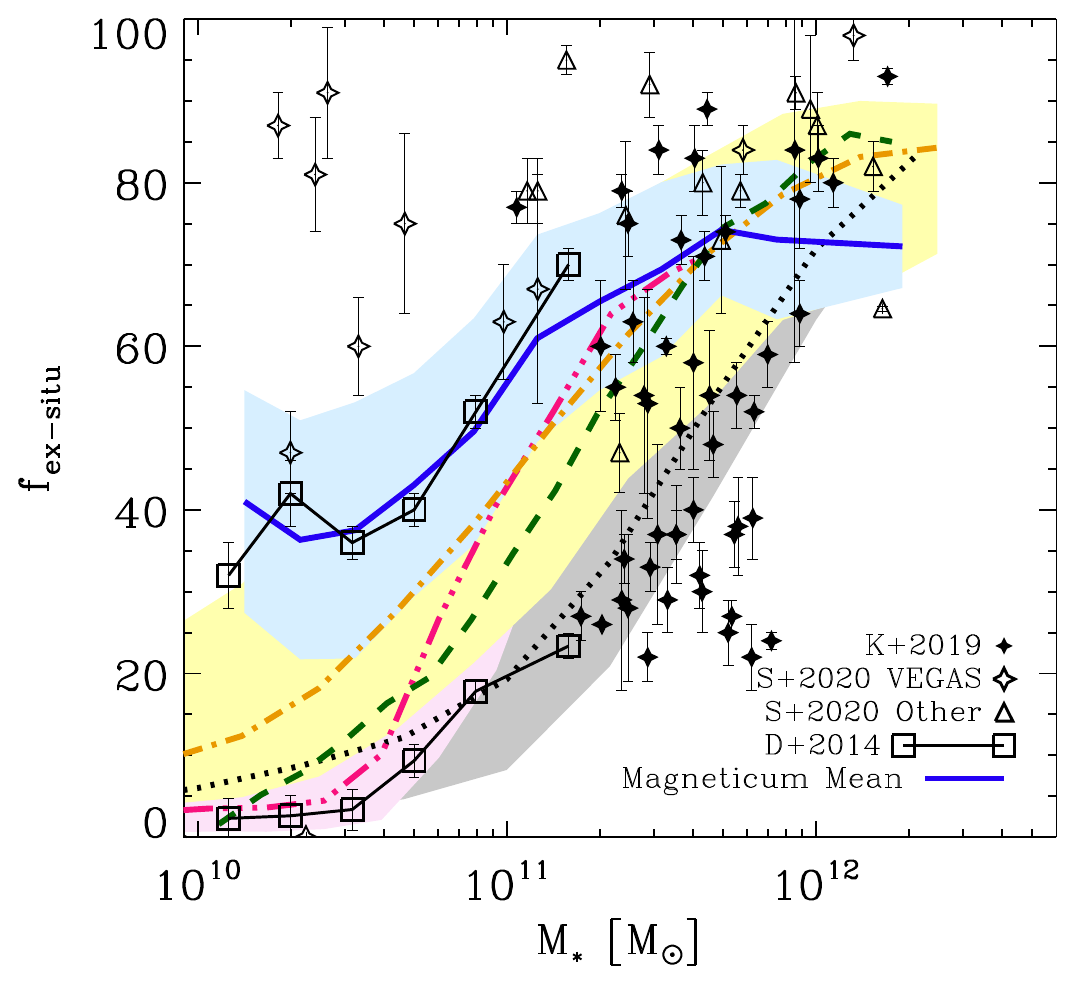}
    \caption{Mass fraction of accreted stars versus stellar mass (same as the upper right panel of Fig.~\ref{fig:fexsitu_comp}) but with observations included as black symbols: Data points from the VEGAS survey \citet{spavone:2020} as open diamonds, with additional data points from the literature as presented by \citet{spavone:2020} as open triangles. Data points for BCGs from \citet{kluge:2019} are shown as filled black diamonds. The values for the stacked SDSS galaxies by \citet{dsouza:2014} are shown as open squares, with the upper line the one obtained for the ETG-like galaxies where a double S\'ersic fit is a good description, and the lower line for the LTG-like galaxies where a third S\'ersic fit is required. The blue solid line and shaded region indicate the Magneticum average values from this work, and all other coloured lines and shaded areas indicate other cosmological simulations as described in Fig.~\ref{fig:fexsitu_comp}.
        }
    \label{fig:fexsitu_obs}
\end{figure}
In Fig.~\ref{fig:fexsitu_obs} we reproduce the top right panel of Fig.~\ref{fig:fexsitu_comp}, showing the ex-situ accretion fraction versus stellar mass. Again, the lines and shading show the results of various models, colors as indicated in Fig.~\ref{fig:fexsitu_comp}, with the blue solid line showing the average fraction of accreted mass for Magneticum galaxies. This time, we included in Fig.~\ref{fig:fexsitu_obs}  observational data from the literature. 
These observations do not directly measure the fraction of accreted mass but rather fit double S\'ersic profiles to the surface brightness profiles of early-type galaxies, as described in Sec.~\ref{sec:42}, inferring the mass fraction from the outer S\'ersic component. 
\begin{figure*}
    \includegraphics[width=0.9\textwidth]{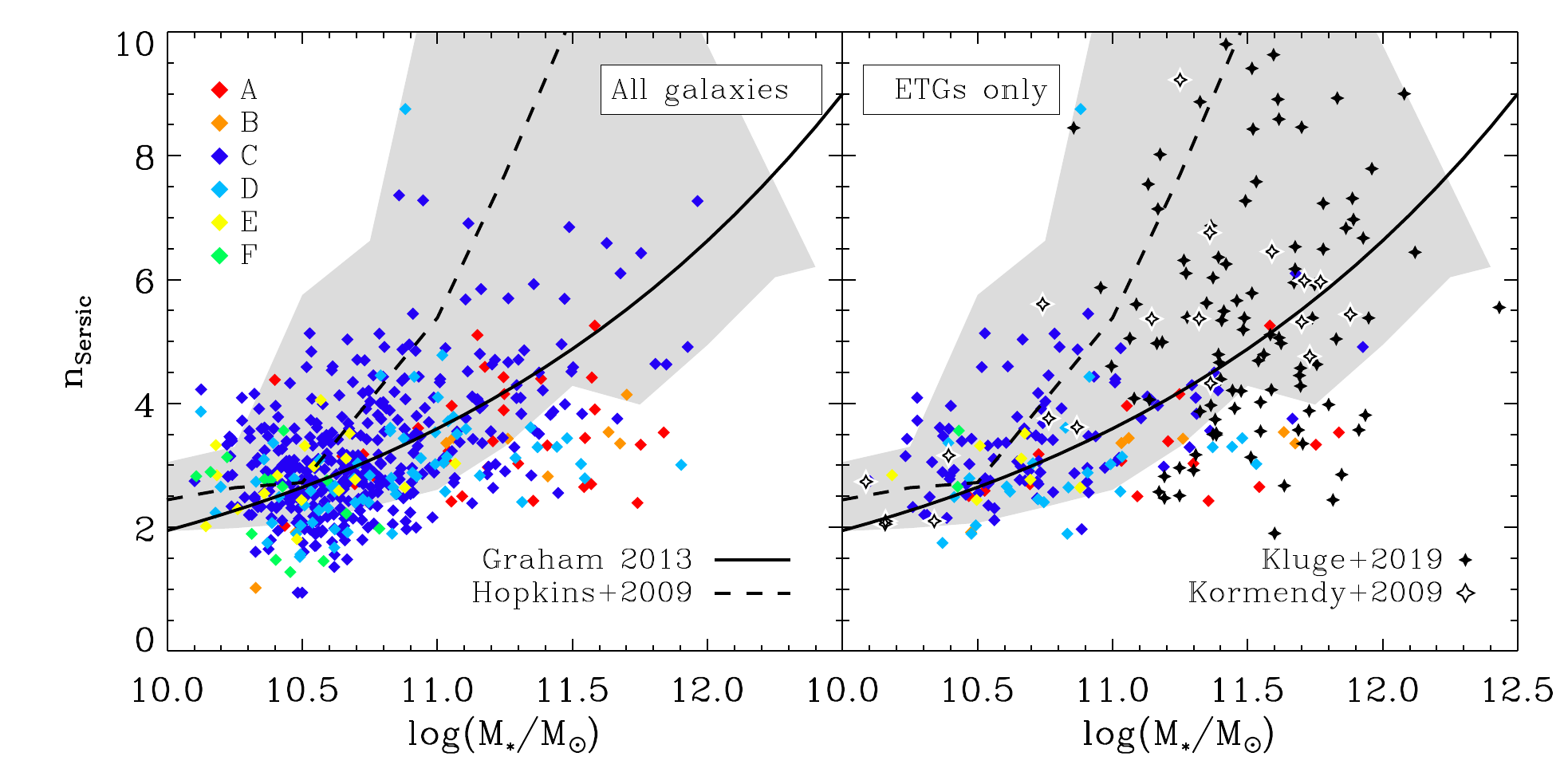}
    \caption{S\'ersic index $n$ versus stellar mass for single S\'ersic fits to the 2D projected mass density profiles of all Magneticum galaxies.
    The dashed black line show the results from simulations by \citet{hopkins:2009}, with the shaded area marking the scatter inferred from their simulations.
    The solid black line is the mean relation from observations by \citep{graham:2013} converted into stellar mass assuming $M/L_\mathrm{B} = 10$.
    \textit{Left panel:} All Magneticum galaxies are coloured according to their profile classes as indicated in the legend. 
    \textit{Right panel:} Only the ETGs from the Magneticum galaxy sample are shown. In addition, open black diamonds show observations of ETGs from \citet{kormendy:2009}, and the solid black diamonds show BCGs from the observations by \citet{kluge:2019}.
              	}
    \label{fig:singlesersic}
\end{figure*}

While all simulations suggest a general trend of increasing accretion fraction for higher mass galaxies, there is no clear trend for the observational proxy (the outer component mass) to vary with stellar mass for the Fornax galaxy sample by \citet{spavone:2020} or the BCG sample by \citet{kluge:2019}. Only the stacked sample provided by \citet{dsouza:2014} shows a decrease in accreted fraction with mass for their ETG sample. In fact, the different observations differ strongly from each others and do not show a clear picture of the accreted fraction estimated through the outer S\'ersic fit component being correlated with the total stellar mass of a galaxy. 
This further supports our results from Sec.~\ref{sec:41}, that dips in the observed surface brightness profile do not, in general, correspond to the true transition from in-situ to accreted dominated material. An alternative approach to estimating accretion fraction may come from star formation histories \citep{2020MNRAS.491..823B} or 2D chemo-dynamical analysis \citep{2019MNRAS.487.3776P}, and needs to be investigated in future studies.

\subsection{Indices from Single and Double S\'ersic Fits}
As a final test, we compare the S\'ersic indices $n$ from the single and double S\'ersic fits to our projected simulated galaxies with observations to ensure that the simulated and observed galaxy samples really are comparable in their radial properties. 
Fig.~\ref{fig:singlesersic} shows the S\'ersic indices $n$ from single S\'ersic fits to the 2D projected mass density profiles of the Magneticum galaxies, for all galaxies in the left panel and ETGs-only in the right panel.
We also include the observed relation for galaxies using eq. 2.7 from \citet{graham:2013} and simply assuming $M/L_\mathrm{B} = 10$ for all galaxies as solid black line.
We find a trend of increasing S\'ersic index $n$ for higher mass galaxies that matches the observed mean trend well, clearly showing that the overall mass distribution of the simulated galaxies is in good agreement with observations.
Galaxies of the different profile classes are well spread in S\'ersic index for a given stellar mass, with no clear trends apart from the fact that the largest S\'ersic indices are clearly found in class~C galaxies.
We also include the range of S\'ersic indices from the empirical model by \citet{hopkins:2009}, which cover a similar mass range than our simulated galaxies. The model predicts on average S\'ersic indices that are, at a given stellar mass, somewhat larger than the Magneticum and the observed galaxies of \citet{graham:2013}, especially at the high mass end, but the scatter range is very similar, and the overall trend of larger S\'rsic indices with larger stellar mass is in good agreement.

When limiting our galaxy sample to ETGs-only (right panel of Fig.~\ref{fig:singlesersic}), we find the S\'ersic indices to be slightly larger on average. We additionally include the observational data for individual ETGs from \citet{kormendy:2009} and \citet{kluge:2019}, further showing that the simulations also provide good descriptions of the stellar mass distributions of ETGs especially. There is a slight tendency for galaxies with stellar masses above $\log(M_*)>11.5$ to have larger S\'ersic indices in the observed samples than in our simulated sample, however, at this mass range our simulated sample is statistically not representative anymore, especially since there are only 4 BCGs in our sample while the observations by \citet{kluge:2019} are BCGs only\footnote{For a more detailed comparison of the BCG properties from a larger Magneticum simulation volume with observations, see Remus et al., in prep.}.
\begin{figure*}
    \includegraphics[width=0.9\textwidth]{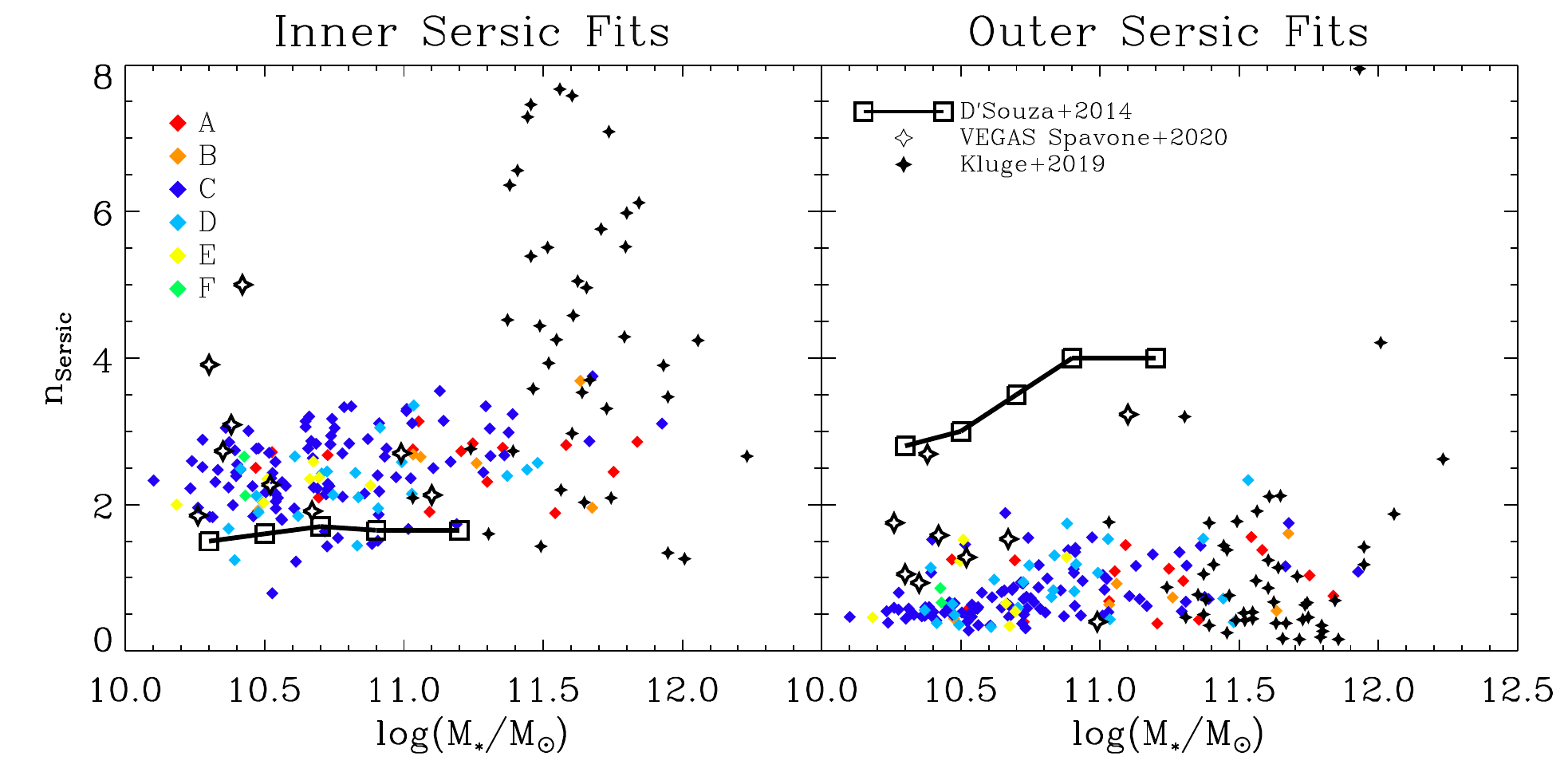}
    \caption{S\'ersic index $n$ versus stellar mass for double S\'ersic fits to the 2D projected mass density profiles of the Magneticum galaxies, with colours as indicated in the legend.
    Observations from the VEGAS survey \citep{spavone:2020} are shown as open black diamonds, and BCGs from \citet{kluge:2019} as solid black diamonds.
    The black lines mark the values obtained by \citet{dsouza:2014} from fits to over $45,500$ galaxies.  
    \textit{Left panel:} Inner S\'ersic fits. \textit{Right panel:} Outer S\'ersic fits.
    	}
    \label{fig:doublesersic}
\end{figure*}

For double S\'ersic fits, we here focus only on ETGs as the observational comparison samples only include ETGs. As shown in Fig.~\ref{fig:doublesersic}, we find the inner profiles of our simulated ETGs to have S\'ersic indices of $n \sim 2.5$ on average (left panel), and the outer profiles to have S\'ersic indices of $n\sim{}$0.5--1, on average (right panel).
We find little variation of these indices with stellar mass, if all there is a slight average trend for the inner S\'ersic indices of more massive galaxies to be slightly larger.
The agreement with the 9 ETGs from the VEGAS survey \citep{spavone:2020}, with masses $>10^{10} M_\odot$, is reasonable, with a slight tendency for the obersved outer S\'ersic indices to be larger than for the simulated ETGs.
Compared to the observations of BCGs by \citet{kluge:2019}, we find that the agreement with their outer S\'ersic indices is rather good, while their inner S\'ersic are on average larger than those of our simulated galaxies. However, this could also be due to the fact that most of the Magneticum ETGs in the mass range comparable to the sample by \citet{kluge:2019} are not BCGs.

Additionally, we also show in Fig.~\ref{fig:doublesersic} the inner and outer S\'ersic indices from the stacked observations by \citet{dsouza:2014}, using their double S\'ersic fits (solid lines).
As can clearly be seen they differ rather strongly from the simulated galaxies, but also the other observations, with their inner slopes generally smaller and their outer slopes much larger in comparison.
We note that \citet{dsouza:2014} also fit triple S\'ersic profiles to their highest mass galaxies.
In this case, their outer S\'ersic fits have $n\sim 1.5$, which is much closer to our simulation values and the other observations, indicating that their inner slopes are actually really ``inner'' slopes which we do not fit in this work to avoid resolution issues. This clearly highlights the importance of clear definitions regarding the fitted regions of galaxies when performing comparisons.

Overall, we find reasonable agreement between the S\'ersic indices $n$ predicted by our simulated galaxies and the observed S\'ersic indices, especially of ETGs, for both single and double S\'ersic fit indices, clearly showing that the simulated galaxies used in this work capture the observed matter distributions. 
This clearly shows that our result, that double S\'ersic fits do \textit{not} describe the in-situ and accreted components of galaxies, is applicable to observations. We rather suggest that the double S\'ersic fit (excluding the central inner bulge areas) describes the relaxed inner and the unrelaxed outer stellar (halo) components of galaxies and could therefore be used to distinguish the outer stellar halo from a galaxy.

\section{Summary and Conclusions}\label{sec:5}
Using the Magneticum simulation we have studied a sample of 511 model galaxies in the log mass range $10.3 < M_* < 12$. These simulated galaxies reproduce well the observed galaxy size-mass relation. We also find that the fraction of accreted material, as a function of total halo and stellar mass, reveals similar trends to those found by previous simulations. One notable difference is that our simulations predict somewhat higher accretion fractions in the lower mass galaxies. 

We examined the stellar mass density profiles of our sample, split in accreted and in-situ components, and classified them into 6 classes depending on the profile type: The most common class reveals, at intermediate radii, a transition radius where the accreted and the in-situ component are equal, with the in-situ component being dominant in the centre and the accreted component dominating the outskirts. This class is comparable to the profiles found in previous studies.
However, for about 30\% of our galaxies we find different profiles: Some are accretion dominated at all radii, even in the centre; another group of galaxies is in-situ dominated at all radii; and most interestingly, we find one class of galaxies that have even \textit{two} transition radii at which the in-situ and accreted material are equal, with an accretion dominated core, an in-situ dominated shell around it, and an accretion dominated outskirt. This is actually the second most common class of galaxies in our sample.

We show that these profile classes correlate with galaxy mass, and that the type of mergers they undergo help to shape their profiles. We especially show that the amount of gas that is involved in these mergers is more important in shaping these profiles than the actual merger, and that the most common class is to about 70\% \textit{not} dominated by major mergers but rather smaller merger events. Their outer regions are largely built up by dry minor and mini mergers, clearly showing the importance of minor and especially mini mergers in shaping the outer stellar halos of galaxies.

We find that galaxies with high in-situ fractions (low accretion fractions) tend to be lower mass galaxies with smaller halfmass radii, and we see a weak trend for high in-situ fraction galaxies to have lower central dark matter fractions, with the exception of the overall in-situ dominated galaxies that have clearly larger central dark matter fractions at a given stellar mass than galaxies of the most common class, typical for what is found in disk galaxies. 
We measure the radius between the in-situ and accretion-dominated regions for those galaxies that reveal a clear transition, which are the majority of our sample. This transition radius is found to weakly be inversely correlated with stellar mass, but strongly correlated with the in-situ fraction for our most common class of galaxies. However, galaxies of the other classes that have one or even two transition radii, do not follow the same relations.

In order to compare our 3D profiles with observations, we projected the stellar mass density in different angles for each galaxy to be able to directly compare with observed surface brightness profiles. We find that the transition radius from in-situ to accretion dominated profiles seen in many 3D profiles also occurs in all projections, but always at different radii in the 2D profiles, usually at smaller radii. None of our galaxies changes its in-situ-profile class during projections.
We also find that, similar to observations, our projected stellar mass surface density profiles usually require a double S\'ersic fit to be described accurately, with S\'ersic indices for both components similar to the range of observed S\'ersic indices. 
However, we clearly see that these two S\'ersic components usually do \textit{not} describe the underlying in-situ and accreted components, but are rather disjunct from those. Only in very few cases the crossing radius of the two S\'ersic components coincides with the transition radius of the galaxy. 
Even worse, we also clearly see that most galaxies that are dominated by accreted stars at all radii, still require a double S\'ersic fit to describe the stellar surface density profiles, thus having a crossing radius but no transition radius.

In other words, we clearly conclude that the dip seen in 2D profiles does {\it not} correspond to the true transition radius between in-situ and accretion dominated regions.
Similarly, any mass inferred from these double-S\'ersci fits will {\it not} trace the true in-situ or accreted mass of a galaxy. 
Thus, fits to the dips seen in some observed surface brightness profiles of early-type galaxies are {\it not} a true measure of a galaxy's accretion material. 
However, they do hold some information about the assembly history of that galaxy, as we find indications that these dips are more likely an indication for the transition from the inner (in-situ and massive merger dominated) core of a galaxy to its stellar halo, mostly accreted through minor and mini mergers, similar to the ICL component around BCGs. To confirm this, a more detailed study including also the radial kinematics of a galaxy in addition to its density component is needed in the future, to disentangle the formation pathways of galaxies from observational tracers.  

\section*{Acknowledgements}
We thank Klaus Dolag and Felix Schulze for very useful discussions.
We also thank Thomas Davison, Enrica Iodice, and Marilena Spavone for their helpful comments. 
We also acknowledge funding from the DAAD PPP Germany-Australia Exchange Program.
The Magneticum Pathfinder simulations were partially performed at the Leibniz-Rechenzentrum with CPU time assigned to the Project ``pr86re'', supported by the DFG Cluster of Excellence ``Origin and Structure of the Universe''. We are especially grateful for the support by M. Petkova through the Computational Center for Particle and Astrophysics (C2PAP).

\section*{Data Availability}
The data underlying this article will be shared on reasonable request to the corresponding author.




\bibliographystyle{mnras}
\bibliography{bibliography} 



\appendix

\bsp	
\label{lastpage}
\end{document}